\documentclass[twocolumn]{aastex62}

\usepackage{amssymb}
\usepackage{amsmath}

\DeclareMathOperator{\sech}{sech}

\def\beq{\begin{equation}\begin{aligned}}
\def\eeq{\end{aligned}\end{equation}}

\newcommand{\Like}{\mathcal{L}}

\begin{document}

 \title{Stellar Profile Independent Determination of the Dark Matter Distribution of the Fornax Local Group Dwarf Spheroidal Galaxy} 

\author{Sasha R. Brownsberger}
\affiliation{Department of Physics Harvard University, Cambridge, MA 02138 , USA}

\author{Lisa Randall}
\affiliation{Department of Physics Harvard University, Cambridge, MA 02138 , USA}

\correspondingauthor{Sasha Brownsberger}
\email{sashabrownsberger@g.harvard.edu}

\shortauthors{Brownsberger \& Randall 2019}

\nocollaboration

\begin{abstract}
The local group dwarf spheroidal galaxies (LG dSphs) are among the most promising astrophysical targets for probing the small scale structure of dark matter (DM) subhalos.  We describe a method for testing the correspondence between proposed DM halo models and observations of stellar populations within LG dSphs.  By leveraging the gravitational potential of any proposed DM model and the available stellar kinematical data, we can derive a prediction for the observed stellar surface density of an LG dSph that can be directly compared with observations.  Because we do not make any reference to an assumed surface brightness profile, our model can be applied to exotic DM distributions that produce atypical stellar density distributions.  We use our methodology to determine that the DM halo of the Fornax LG dSph is more likely cored than cusped, ascertain that it is characterized by a semi-minor to semi-major axis ratio in minor tension with simulations, and find no substantial evidence of a disk within the dSph's larger DM halo. 
\end{abstract}


\section{Introduction and Background} 
For many decades, evidence pointing to the existence of some gravitationally active and electromagnetically unobservable (or at least hitherto unobserved) material has accrued from a variety of astronomical and cosmological sources \citep{Rubin, Bennett93, Corbelli, Jones06, Clowe, Hinshaw13, Planck16, Salucci18}.  Today it is generally accepted that some unknown dark matter (DM) with properties not described by the standard model of particle physics is primarily responsible for these gravitational anomalies \citep{Peter12, Zurek14, Lisanti15, Majumdar15}.  

Despite the impressive experimental and theoretical work of many researchers \citep{Undagoitia16, Liu17, Scherrer86, Baer16} the true nature of DM remains a mystery.  
Nonetheless, the $\Lambda$CDM cosmological paradigm of a flat universe dominated today by a cosmological constant ($\Lambda$) and by nonrelativistic, collisionless (`cold') dark matter (CDM) is in general agreement with a diverse set of cosmological probes, including the power spectrum of the Cosmic Microwave Background (CMB) anisotropies \citep{Bennett93, Jones06, Hinshaw13, Planck16}, the measured flux of type Ia supernovae \citep{Riess98, Perlmutter99, Barris04, Hicken09, Kessler09, Sullivan11, Takanashi17, Foley18, Scolnic18}, the baryonic acoustic oscillations \citep{Eisenstein05, Percival10, desBourboux17, Slepian17, Bautista17, Beutler17, Ross17, Ata18}, and the cosmological history of structure formation \citep{Davis85, Moore99, Garrison-Kimmel14, Schaller15}.  

However, an inconsistency between the predictions of the $\Lambda$CDM paradigm and astronomical observations remains unresolved: $\Lambda$CDM simulations predict DM halos with centrally peaked (`cusped') mass density profiles \citep{Navarro97}, but observations of galaxies and dwarf galaxies identify DM halos with centrally flat (`cored') interior mass density profiles \citep{Moore94, Flores94, Burkert95, MooreandQuinn99, Navarro00, deBlok02, Alam02}.  This inconsistency between simulations and predictions is known as the `core-cusp' problem.

The `core-cusp' problem may evince DM properties beyond the CDM paradigm.  \cite{McDonald09, Angulo13, Destri14, Gonzalez-Samaniego16} argue that DM particles would be less prone to forming cusped halos if primordial DM either remained relativistic for longer or decoupled later than canonically predicted.  \cite{Rocha13, Peter13, Robles17, Sokolenko18, Harvey18, Robertson19, Sameie18, Fitts19, Robles19} claim that self-interacting dark matter (SIDM) could virialize dense DM cusps, turning them into a cores.  

The inclusion of baryonic effects (generally ignored for computational ease) could resolve the `core-cusp' problem without the need for physics beyond the $\Lambda$CDM paradigm.
Via various methods, including highly energetic supernovae and stochastic oscillations in local density, baryons can transfer energy to DM, turning a galaxy's DM halo from cusped to cored \citep{Navarro96, Read05, Governato10, Pontzen12, Governato12, Maccio12, Teyssier13, Ogiya14, El-Zant16, Allaert17, Dashyan18, Popolo18}.
In this scenario, an isolated DM halo with sufficiently little stellar mass would receive an insufficient injection of energy from this baryonic feedback to turn a cusp into a core.  The precise definition of `sufficiently little stellar mass' remains a point of contention amongst \cite{Penarrubia12, GarrisonKimmel13, DiCintio14b, Onorbe15, Lucia18, Dutton19}.

Determining if the `core-cusp' problem can be solved within the $\Lambda$CDM paradigm necessitates ever more accurate measurements of existing DM structures.  The ideal systems for studying the distributions of DM at small scales are those that are DM dominated and that also contain enough baryons to form observable gravitational tracers.  The Local Group dwarf spheroidal galaxies (LG dSphs) constitute a population of such systems. 

LG dSphs, summarized as a population by \cite{Mateo98, Penarrubia08, McConnachie12, Munoz18}, are some of the most DM dominated objects yet identified, and many have stellar populations characterized \citep{Mateo97, Kleyna02, Tolstoy04, Coleman05, Munoz05, Koch06, Battaglia06, Koch07, Martin07, Simon07, Koch08, Battaglia08, Walker09, Walker09_data, Kirby10, Ural10, Battaglia11, Majewski13, Hendricks14, Walker15, Caldwell17, Kacharov17, Gonzalez18, Conselice18}.  They are also generally cold enough for their stars to serve as good tracers of their underlying gravitational potentials.  Due to this combination of DM domination and observability, LG dSphs may contain observable signatures of DM substructure largely unperturbed by the effects of baryonic physics.  

\begin{deluxetable*} { C C C C C} 
 \tablecaption{Summary of some previous efforts to determine the internal DM distribution of LG dSphs.  \label{tab:prevEfforts}}
\tablehead  {\colhead{Reference}  & \colhead{dSph(s) studied} & \colhead{Brief summary of method$^ {*}$} & \colhead{Some assumptions$^ {\dagger, \ddagger}$ } & \colhead{Cored or cusped?$^{\S}$}                  }               
\startdata
\shortstack{\cite{Lokas02} \\ \textcolor{white}A }               & \shortstack{Draco \\ Fornax }                                     & \shortstack[l]{Jeans analysis of DM halos and the \\ \ \ the moments of stellar velocities. }      & \shortstack{SS, E, ASP \\ \textcolor{white} A }             & \shortstack[l]{Cored \\ \textcolor{white} A }         \\
\shortstack{\cite{Kleyna03} \\ \textcolor{white}A }               & \shortstack{Ursa Minor \\ \textcolor{white}A}                                     & \shortstack[l]{N-body simulation of the stability \ \\ \ \  of stellar clumps.}      & \shortstack{NB, STP, PO \\ SS}             & \shortstack[l]{Cored \\ \textcolor{white} A}         \\
\shortstack{\cite{SanchezSalcedo06}   }              & \shortstack{Fornax }                                    & \shortstack[l]{Analysis globular cluster evolution.}                                & \shortstack{STP, PO, SS }     & \shortstack[l]{Cored }             \\   
\shortstack{\cite{Goerdt06}   }                   &  \shortstack{Fornax}                                     &  \shortstack[l]{Analysis globular cluster evolution.}                         & \shortstack{NB, PO, SS}       &\shortstack[l]{Cored }              \\  
\shortstack{\cite{Penarrubia10}  \\ \textcolor{white}A}                & \shortstack{All LG dSphs \\ \textcolor{white}A }                                  & \shortstack[l]{Analysis of the robustness of dSph \ \\ \ \ structure under tidal stripping.}                                   & \shortstack{NB, PIO, SS \\ \textcolor{white} A}    & \shortstack[l]{Cusped \\ \textcolor{white} A}             \\   
\shortstack{\cite{Walker11}  \\ \textcolor{white}A   \\ \textcolor{white}A \\ \textcolor{white} A}                  &  \shortstack{Fornax \\ Sculptor   \\ \textcolor{white}A  \\ \textcolor{white}A }                                     & \shortstack[l]{Measurement of the DM gradient \ \ \\ \ \ from Jeans analysis of multiple \\ \ \ RG populations.}                              & \shortstack{E, SS, ASP, \\ NSG, IVD, CVD \\  \textcolor{white} A \\ \textcolor{white} A}    & \shortstack[l]{Cored \\ \textcolor{white} A \\ \textcolor{white} A \\ \textcolor{white} A }              \\   
\shortstack{\cite{Agnello12} \\ \textcolor{white}A }                  & \shortstack{Sculptor \\ \textcolor{white}A}                                    &  \shortstack[l]{Virial analysis of distinct stellar \ \ \ \ \\ \ \   populations. }      &                              \shortstack{NSG, ASP  \\ \textcolor{white} A}  & \shortstack[l]{Cored \\ \textcolor{white} A}               \\   
\shortstack{\cite{Hayashi12} \\ \textcolor{white}A \\ \textcolor{white}A  }           & \shortstack{Carina, Draco \\ Fornax, Leo I \\ Sculptor, Sextans}                                      & \shortstack[l]{Least-squares fitting of \ \ \ \ \ \ \ \   \ \ \ \ \ \ \  \\ \ \ observations with cylindrical \\ \ \  Jeans analysis. \ \ \ \   }                               &  \shortstack{E, SS, ASP, \\ NSG \\ \textcolor{white}A   }         &  \shortstack[l]{Either \\ \textcolor{white} A \\ \textcolor{white}A}                     \\
\shortstack{\cite{Lora13}  \\ \textcolor{white}A }               & \shortstack{Sextans \\ \textcolor{white}A}                                     &  \shortstack[l]{N-body simulation of the stability \ \\ \ \  of stellar clumps.}                               & \shortstack{NB, STP,  SS, \\ NSG, IVD}     &  \shortstack{Cored \\ \textcolor{white} A}            \\   
\shortstack{\cite{Breddels13} \\ \textcolor{white}A \\ \textcolor{white}A   }          & \shortstack{Sculptor \\ \textcolor{white}A \\ \textcolor{white}A  }                                     & \shortstack[l]{Measurement of enclosed mass \ \ \ \ \ \ \\ \ \ and velocity dispersion with \\ \ \ orbit superposition. }                               &  \shortstack{SS \\ \textcolor{white}A \\ \textcolor{white}A }     & \shortstack{Cored \\ \textcolor{white} A  \\ \textcolor{white}A }                \\   
\shortstack{\cite{Majewski13}  \\ \textcolor{white}A  }          &  \shortstack{Sagittarius  \\ \textcolor{white}A }                                     & \shortstack[l]{Jeans analysis of a newly \ \  \ \ \ \ \  \ \  \ \ \ \ \\ \ \ discovered, cold population. }                              & \shortstack{E,  SS,  ASP, \\ NSG,  CVD   }    & \shortstack{Either \\ \textcolor{white} A}          \\ 
\shortstack{\cite{Richardson14} \\ \textcolor{white}A \\ \textcolor{white}A }      & \shortstack{Sculptor \\ \textcolor{white}A \\ \textcolor{white}A  }                                     & \shortstack[l]{Measurement of newly-defined \ \ \ \ \ \ \\ \ \  dSph shape parameters to \\ \ \ constrain DM profiles. }                   & \shortstack{E,  SS,  ASP, \\ NSG \\ \textcolor{white}A}       & \shortstack{Cusped \\ \textcolor{white} A  \\ \textcolor{white}A}                         \\
\shortstack{\cite{Mamon15}  \\ \textcolor{white}A   }           & \shortstack{Fornax, M87 \\ \textcolor{white}A  }                                     & \shortstack[l]{Use of MAMPOSSt technique to  \\ \ \  study gravitational tracers. }                               &  \shortstack{SS, ASP \\ \textcolor{white}A }     & \shortstack{Cusped for M87 \\ Unclear for Fornax  }                       \\
\shortstack{\cite{Pace16}  \\ \textcolor{white}J \\ \textcolor{white}R }                  &  \shortstack{Ursa Minor \\ \textcolor{white}A \\ \textcolor{white}A }                                     & \shortstack[l]{Measurement of the DM gradient \ \ \\ \ \ from Jeans analysis of multiple \\ \ \ RG populations.}                              & \shortstack{E, SS, ASP, \\ NSG, IVD, CVD  \\ \textcolor{white}A}    & \shortstack{Cored \\ \textcolor{white} A  \\ \textcolor{white}A}               \\  
\shortstack{\cite{Contenta17} \\ \textcolor{white}A }                   & \shortstack{Eridanus II \\ \textcolor{white}A}                                     & \shortstack[l]{N-body simulation of the stability \ \\ \ \  of stellar clumps.}                              &  \shortstack{SS, NB, STP, \\ PO, NSG}     & \shortstack{Cored \\ \textcolor{white} A}               \\      
\shortstack{\cite{Caldwell17} \\ \textcolor{white}A  \\ \textcolor{white}A }                   &  \shortstack{Crater II \\ \textcolor{white}A  \\ \textcolor{white}A}                                     & \shortstack[l]{Measurement of the DM gradient \ \ \\ \ \ from Jeans analysis of multiple \\ \ \ RG populations.}                              & \shortstack{E, SS, ASP, \\ NSG, IVD, CVD  \\ \textcolor{white}A}  &\shortstack{Cored \\ \textcolor{white} A}              \\ 
\shortstack{ \cite{Strigari17} \\ \textcolor{white}A  \\ \textcolor{white}A   }           & \shortstack{Sculptor \\ \textcolor{white}A  \\ \textcolor{white}A   }                                     & \shortstack[l]{Matching observations to general \ \ \\ \ \  parametric forms of energy and  \\ \ \ angular momentum. }                               & \shortstack{E, SS, ASP, \\ NSG  \\ \textcolor{white}A  }       &   \shortstack{Either \\ \textcolor{white} A \\ \textcolor{white}A  }    \\ 
\shortstack{ \cite{Boldrini18} \\ \textcolor{white}A    }           & \shortstack{Fornax\\ \textcolor{white}A    }                                     & \shortstack[l]{ N-body simulation, accounting for \\ \ \  dynamic effects.}                               & \shortstack{NB  \\ \textcolor{white}A  }       &   \shortstack{Cored \\ \textcolor{white} A \\ \textcolor{white}A  }    \\ 
\shortstack{ \cite{Read18} \\ \textcolor{white}A  \\ \textcolor{white}A    }           & \shortstack{Draco \\ \textcolor{white}A  \\ \textcolor{white}A }                                     & \shortstack[l]{Numerical fitting of Jeans \ \ \ \ \ \ \ \ \ \ \  \\ \ \ equation to morphological and  \\ \ \ kinematical data. }                               & \shortstack{E, SS  \\ \textcolor{white}A  \\ \textcolor{white}A   }       &   \shortstack{Cusped\\ \textcolor{white} A \\ \textcolor{white}A   }         \\ 
\shortstack{ \cite{Moskowitz19} \\ \textcolor{white}A  \\ \textcolor{white}A  \\ \textcolor{white}A   }           & \shortstack{Fornax,
Leo I \\ Leo II,  \\ Reticulum II, \\ and others  }                                     & \shortstack[l]{Fitting observed stellar positions \  \\ \ \ to generalized stellar profile \\ \ \ consisting of Plummer models  \\ \ \ validated using mock data.  }                               & \shortstack{PO, SS  \\ \textcolor{white}A  \\ \textcolor{white}A  \\ \textcolor{white}A  }       &   \shortstack{Cored\\ \textcolor{white} A \\ \textcolor{white}A  \\ \textcolor{white}A  }                              
 \enddata
 \tablenotetext{*}{The clever and sophisticated analyses of these works cannot be satisfactorily described with a single sentence.  We encourage the interested reader to explore the original publications. }
 \tablenotetext{\dagger}{The abbreviations used to denote commonly made assumptions are as follows: NB - dSph evolution described by N-body analyses, STP - conclusions drawn from a relatively small tracer population, PO - conclusions drawn only using measurements of tracer positions, SS - spherical symmetry, CS - spheroidal symmetry, ASP - archetypal stellar profile, E - equilibrium, NSG - negligible stellar gravity, IVD - isotropic velocity dispersion, CVD - constant velocity dispersion.  }
 \tablenotetext{\ddagger}{Some techniques may suffer limitations or make assumptions beyond those listed.}
 \tablenotetext{\S}{Different authors conclude that certain DM halos are more likely cored or cusped with varying degrees of certainty.}
 \tablecomments{\cite{Laporte13, Gonzalez-Samaniego17} contend that the fundamental triaxiality of dSphs does not significantly affect the conclusions of \cite{Walker11, Agnello12, Pace16, Caldwell17}.}
\end{deluxetable*}

Studying LG dSphs is complicated by the difficulties inherent in characterizing the orbits of pressure-supported stellar systems and in measuring the velocities of extragalactic stars.  Over the years, a variety of methodologies for probing the DM substructure of LG dSphs have been implemented, several of which we summarize in Table \ref{tab:prevEfforts}.  No particular methodology has been generally accepted as ideal.  
 All of these efforts have their own particular advantages, but the list of questions that they can answer is inherently limited by the assumptions that they make.  
Particularly, many techniques assume that the dSphs' stellar distributions adhere to archetypal stellar profiles, such as the Plummer profile of \cite{Plummer11}.  Such assumptions could preclude the identification of atypical stellar structure and, as \cite{Moskowitz19} note, may undermine attempts to characterize underlying DM halo profiles. 

In this work, we describe a new analysis method, somewhat similar to those of \cite{Read18, Moskowitz19}, that allows the expected positions of stellar tracers to be determined from a proposed DM distribution and kinematical model without an assumed stellar profile. 
This method has enabled us and will enable other astronomers to search for evidence of unusual DM substructures unadulterated by assumed stellar distributions.
We view this method as supplementary to previous analyses.

We use this method to answer the following questions about the Fornax LG dSph:
\begin{enumerate}
\item For various DM halo models, what are the parameters that best match the structure of the dSph? 
\item Are any of these parameters in tension with simulations? 
\item Is the dSph described better by a prolate or an oblate DM halo?  
\item Is the dSph described better by a cored or a cusped DM halo? 
\item Is there a disk-like structure sitting inside of the dSph's larger DM halo?  
\end{enumerate}
In brief, our answers to these questions are as follows: 
\begin{enumerate}
\item See Table \ref{tab:fornBestFitWOD}.  
\item Slightly.  The dSph's best fit DM halo is more spheroidal than about eighty percent of DM halos predicted by $\Lambda$CDM simulations.  
\item No preference.  Both prolate and oblate halos match the dSph data similarly well.  
\item A cored halo.  
\item None larger than one percent of the total halo mass.  
\end{enumerate} 

This work is organized as follows.  
In Section \ref{sec:create}, we use the Jeans equations to determine the three dimensional probability distribution, $\nu$, for a stellar population given an underlying gravitational potential and stellar kinematics.  In Section \ref{sec:compSurfProb}, we derive the likelihood of observing a star at a given position on the sky for a given DM halo by integrating $\nu$ along the line of sight and correcting for inhomogeneities in the observation sensitivity.  In Section \ref{sec:compLikelihood}, we describe how this predicted probability of observing a star is compared with observed data to determine the correspondence between a model and observations.  In Section \ref{sec:measureSig}, we use statistical bootstrapping to determine which of a set of proposed models best matches the data.
In Section \ref{sec:FornaxExample}, we apply this technique to the particular case of the Fornax LG dSph.  We describe the data in Section \ref{sec:data}, the DM models considered in Section \ref{sec:DDDM}, the utilized kinematics in Section \ref{sec:fornKinematics}, the creation of the probability surface density of detection in Section \ref{sec:losApp}, and our method for exploring a parameter space to determine the best correspondence between the data and a model in Section \ref{sec:MCMC}.  Our results, which we discuss in detail in Section \ref{sec:resAndDisc}, demonstrate that the DM distribution of Fornax is better described by a cored halo than a cusped halo and exhibits no evidence of a disk larger than one percent of the total halo mass.  We conclude in Section \ref{sec:conclusion} with some remarks on how this technique could be utilized as more data become available. 
 
\section{Methodology} 
\subsection {Description of the Method} \label{sec:theMethod} 
In this section, we describe our general method for probing the substructure of dSph DM halos without assuming a particular stellar distribution.  This method consists of four steps: 
\begin{enumerate}
\item Derive the volume probability density for the dSph from a proposed DM distribution and a kinematical model (Section  \ref{sec:create}). 
\item Acquire an expected observational surface probability density by integrating the volume probability density along the observer's line of sight and correcting for observational sensitivity (Section \ref{sec:compSurfProb}). 
\item Measure the correspondence between an expected observational surface probability density and observational data (Section \ref{sec:compLikelihood}).  
\item Determine which of several proposed models best matches the available data (Section \ref{sec:measureSig}). 
\end{enumerate} 
In Section \ref{sec:FornaxExample}, we apply this method to probe the DM halo of the Fornax LG dSph.  Some of the lengthy derivations have been relegated to the Appendices.  

\subsubsection{Predicting Probability Densities from DM Models } \label{sec:create}
Here, we describe  how we use an assumed DM halo model and an observationally motivated velocity distribution model to determine the probability that a star belonging to the dSph of interest would be observed at a particular position on the sky.  
 
We start with the Jeans equations \citep{BandT} to relate the overall gravitational potential of the dSph, $\Phi$, to the three-dimensional spatial probability density of stars, $\nu$, and velocity field, $\mathbf{v}$, of a stellar population within the dSph:
\begin{equation} \label{eq:genericJeans} 
\nu \frac{\partial \overline{v}_j}{\partial t} + \nu \overline{v}_i \frac{\partial \overline{v}_j}{\partial x_i} = - \frac{\partial \Phi}{\partial x_j} - \frac{\partial (\nu \sigma^2_{ij})}{\partial x_i} \ ,
\end{equation} 
where a bar denotes averaging over all momenta and $\sigma_{ij} \equiv \overline{v_i v_j} - \overline{v}_i\overline{v}_j $ is the velocity dispersion tensor.  If we postulate a gravitational potential and a stellar velocity field, we can use Equation \ref{eq:genericJeans} to derive $\nu$ without assuming a particular stellar morphology. 

Of course, neither the shape of the gravitational potential nor the details of the stellar kinematics are known precisely for any dSph.  In our method, we calculate the former based on the DM model of interest and infer the latter from available kinematical data.  Equation \ref{eq:genericJeans} then yields the spatial probability distribution that would result from a population with the inferred kinematics moving through the hypothesized DM distribution.  

Our method can, in principle, handle any hypothesized gravitational potential and velocity field.  In Appendices \ref{app:axisymPot} and \ref{app:sphericalPot}, we provide two specific examples of deriving analytical expressions for $\nu$ given archetypal gravitational potentials and stellar kinematical models when the dSph is assumed to be in equilibrium.  In Appendices \ref{app:haloPot} and \ref{app:diskPot}, we derive the gravitational potential of several archetypal DM models.  

\subsubsection{Computing Surface Probability Density of Observation} \label{sec:compSurfProb}

To compare a theoretical prediction to astronomical data, we project the theoretical stellar morphology, $\nu$, onto the sky of the observer.  This projected surface probability density, $S$, is defined by
\begin{equation} \label{eq:SDef}
S(x_{\textrm{sky}}, z_{\textrm{sky}}) = \int_{-D} ^ {\infty} d y_{\textrm{sky}} \ \nu (x_{\textrm{sky}}, y_{\textrm{sky}}, z_{\textrm{sky}}) \ ,
\end{equation}
where $D$ is the distance between the observer and the center of the target galaxy.  The relation between the coordinates of the galaxy and those of the observer are determined by the geometry of the system, including the orientation angle of the dSph relative to the observer.  

Computing $S$ from Equation \ref{eq:SDef} in practice usually requires numerical integration along the line of sight for some set of points in the viewing plane, followed by interpolation between those points.  We provide an example of computing $S$ using such a technique as part of our analysis of the Fornax LG dSph in Section \ref{sec:losApp}.  

To compare our theoretically predicted surface probability density to observations of stars, we must account both for the predicted physical distribution of stars and the ability of the observer to detect those stars.  The model-predicted surface probability density that a star is observed at position $(x,z)$, $O(x,z)$, is the normalized product of the surface probability density of a star being located at that position, $S(x,z)$, and of the surface probability density of the observer detecting a star at that position, $D(x,z)$: 
\begin{equation} \label{eq:defObserveProb}
O(x,z) = \frac{S(x,z)D(x,z)}{\int_{\textrm{sky}}dx' dz' \ S(x',z')D(x',z')} \ .
\end{equation}
We refer to $O(x,y)$ as the  `observational surface probability density'.  $D(x,z)$ is determined by the details of the data collection.  

The likelihood, according to the proposed model, of detecting a random star in the population under study in coordinate patch $dA_{\textrm{sky}} \equiv ((x,x+dx), (z,z+dz))$ is 
\beq \label{eq:probOfDetection} 
 P(\textrm{star}  \in dA_{\textrm{sky}}) = O(x,z) dx dz \ .
\eeq
We have derived this observational surface probability density without any assumed stellar profile.  We compare $O(x,y)$ to observations of stars to determine the model's relative goodness of fit.  

\subsubsection{Comparing Observations and Predictions} \label{sec:compLikelihood}
To evaluate the correspondence between am observational surface probability density and a data set, we build on the analysis described in the Appendix of \cite{Richardson11}.

We assume that the studied stellar population contains $N_{\star}$ detected stars and break the sky area of interest into a set of $N_{\textrm{bin}}$ bins of equal area, $A$.  We choose the bins such that bin $i$ is centered at sky coordinate $(x_i,z_i)$ and that $O(x,z)$ is roughly constant over bin $i$.  We also assume that the probability density of observing a particular number of stars in bin $i$ is described by a Poisson distribution.  Under these assumptions, the likelihood, $\Like_i$, of observing $n_i$ stars in bin $i$ is
\beq
\Like_i(n_i | O, N_{\star} ) = \frac{e^{-N_{\star, i} } (N_{\star, i}  ) ^ {n_i }}{n_i !} \ ,
\eeq
where $N_{\star, i}$, the expected number of stars in bin $i$, is given by 
\beq
N_{\star, i}  = N_{\star} \int_{\textrm{bin } i} dx' dz' \ O (x',z')  \simeq  N_\star O(x_i,z_i) A \ .
 \eeq
 The total likelihood, $\Like$, that a considered distribution will give rise to the observed stars at the observed positions is the product of the likelihoods of each bin:
\beq \label{eq:probOfObservationPreSmallBin}
\Like & = \prod_{i=1}^{N_{\textrm{bin}}} \Like_i(n_i | (O, N_{\star})) 
         = \prod_{i=1}^{N_{\textrm{bin}}}  \frac{e^{-N_{\star, i} } (N_{\star, i}  ) ^ {n_i }}{n_i !} \\
         &=  \Big ( e^{-\sum_{i=1}^{N_{\textrm{bin}}} {N_{\star, i} }} \Big ) \Big ( \prod_{i=1}^{N_{\textrm{bin}}}  \frac{ (N_{\star, i}  ) ^ {n_i }}{n_i !} \Big ) \ .
\eeq

We now decrease the bin area until each bin contains zero or one observed stars:
\begin{equation} 
  n_i = 
  \begin{cases} 
      0 & \textrm{no star in bin } i \\
      1 & \textrm{one star in bin } i
   \end{cases} 
   \ .
\end{equation}
Equation \ref{eq:probOfObservationPreSmallBin} can be simplified because the discrete sum can be approximated as an integral:  
\beq
 -\sum_{i=1}^{N_{\textrm{bin}}} {N_{\star, i} } &= -N_{\star}  \sum_{i=1}^{N_{\textrm{bin}}}  ({O(x_i,z_i) A}) \\
&= -N_{\star} \int \limits_{dSph}  O(x_i,z_i) d^2 a = -N_{\star} \ , 
\eeq
and because the product can be rewritten: 
\beq
  \prod_{i=1}^{N_{\textrm{bin}}}  \frac{(N_{\star, i}) ^ {n_i }}{n_i !}  
  &=  
  \prod_{i=1}^{N_{\textrm{bin}}} 
  \begin{cases} 
      1 & \textrm{no star in bin } i \ \\
      N_{\star, i}  & \textrm{one star in bin } i
   \end{cases} \\
   &= \prod_{(x_\star,z_\star)} (N_{\star} O(x_\star,z_\star) A)  \\
   &= (N_{\star} A)^{N_\star} \prod_{(x_\star,z_
   \star)} O(x_\star,z_\star) \ ,
\eeq
where $(x_\star,z_\star)$ are the on-sky positions of the observed stars.  

The likelihood that the considered distribution would give rise to the observed set of stars in this particular binning of the sky is thus
\begin{equation}
\Like = e ^ {-N_\star} (N_{\star} A)^{N_\star} \tilde{\Like} \ , 
\end{equation}
where
\begin{equation}
\tilde{\Like} \equiv  \prod_{(x_\star,z_\star)} O(x_\star,z_\star) \ .
\end{equation}
Note that only $\tilde{\Like}$ depends on the predicted observational surface probability density.  

We express the ratio between the likelihoods of two distinct dSph models, $X$ and $Y$, given one set of dSph observations as 
\beq\label{eq:relLike}
 r_{XY} &\equiv \frac{\Like_X}{\Like_Y} = \exp{\Big ( \ln{\frac{\tilde{\Like}_X }{\tilde{\Like}_Y}}\Big )} \\ 
& = \exp{\Big (\ln \prod_{(x_{\star},z_{\star})} \frac{O_X(x_{\star},z_{\star}) }{O_Y(x_{\star},z_{\star})}\Big )} \\
&= \exp{ \Big( \sum_{(x_{\star},z_{\star})} \ln(O_X(x_{\star},z_{\star})) } \\
& \ \ \ \ \ \ \ \ \ \ \ \ \ \ \  - \sum_{(x_{\star},z_{\star})} \ln(O_Y(x_{\star},z_{\star})) \ \Big)  \ ,
\eeq
where $O_X(x,z)$ and $O_Y(x,z)$ are, the surface probability densities of observation, as defined in Equation \ref{eq:defObserveProb}, for model $X$ and model $Y$, respectively.  Such a computation is readily carried out once the forms of $O_X$ and $O_Y$ are determined. 

The quantity $r_{X,Y}$ defines the relative likelihood that dSph models $X$ and $Y$ would give rise to the observed set of stars.  Assuming uniform Bayesian priors on the models, a value of $r_{X,Y} > 1$ suggests that model $X$ describes the dSph better than model $Y$, and a value of $r_{X,Y} < 1$ suggests the converse.  

In Section \ref{sec:FornaxExample}, we use this measurement of relative likelihoods, in conjunction with a Markov Chain Monte Carlo (MCMC) algorithm, to determine which of several DM halo models describes the Fornax LG dSph. 

\subsubsection{Determining the Statistical Significance of Relative Likelihoods} \label{sec:measureSig}
We use the measured value of $r_{X,Y}$ to gain insight into which of dSph models $X$ and $Y$ best replicates an observed distribution of stars in a dSph.  However, those stars constitute only one of the infinite possible stellar configurations that might result from the dSph's true and undetermined stellar probability density.  It is this underlying stellar probability density that we seek to understand.  
We use statistical bootstrapping \citep{Efron79} to measure the distribution of $r_{X,Y}$ values that could result from the dSph's stellar probability density. 

By randomly selecting $N_{\star}$ observed dSph stars with replacement, we simulate an artificial dSph with $N_{\star}$ constituent stars drawn from the true observational probability distribution.  Remeasuring the value of $r_{X,Y}$ for this new artificial dSph, we acquire an additional measurement of the likelihood ratio between models $X$ and $Y$ drawn from the dSph's observational probability density. 
By repeating this process many times, we sample of the $r_{X,Y}$ distribution for the dSph under study.
Provided our sample size is sufficiently large, we can infer the true distribution of the $r_{X,Y}$ statistic , and thus determine the robustness of our conclusion that model $X$ or model $Y$ better describes the dSph. 

\subsection{Example: DM Models for the Fornax Local Group Dwarf Spheroidal Galaxy} \label{sec:FornaxExample}
The Fornax LG dSph (Fornax henceforth) is both bright and of moderate angular size, with a large number of resolvable stars, some of which are luminous enough to give consistently detectable photometric and spectroscopic signals \citep{Mateo91, Irwin95, Bersier00, Battaglia06, Walker09_data, delPino15, Wang19}.  We were interested both in the internal slope of Fornax's DM halo and also in the possible influence of DM on the formation of the flattened structure of young main sequence stars that \cite{Battaglia06, delPino15, Wang19} detected near Fornax's center.  In particular, we sought to determine if this flattened structure formed in part due to the influence of a flattened DM structure, and if such an effect is detectable in the other stellar populations. 
 We were particularly interested in the possibility of a dark disk of the sort considered by \cite{Fan13, Alexander19}.  We note that searches for a dark disk in the Milky Way \citep{Schutz18} rely on assumptions that the stellar tracers are in equiliibrium, and are not yet fully reliable.  
 The interesting stellar structure in Fornax and the published catalogues of line of sight velocities for many of its constituent stars made Fornax an excellent candidate for the application of our method.  

In studying Fornax, we used the analysis of Appendix \ref{app:axisymPot}.  We worked under the assumption that the portion of Fornax under study could be reasonably modeled as a spheroidally symmetric system with a constant velocity dispersion and single fixed rotation speed.  We discuss the results of our analysis in Section \ref{sec:resAndDisc}. 

\subsubsection{Data Selection and Population Disambiguation for Fornax} \label{sec:data}
We now discuss the details of selecting and calibrating the data used in our study of Fornax.   

Although the flattened stellar structure that first drew our attention to Fornax was identified in a population of young main sequence stars, Fornax's population of Red Giant (RG) stars is the only population for which we had velocity measurements.  Because our analysis required knowledge of a stellar population's kinematics, we could apply our method only to Fornax's RG stars.  

\begin{figure}
    \centering
    \includegraphics[width=0.5 \textwidth]{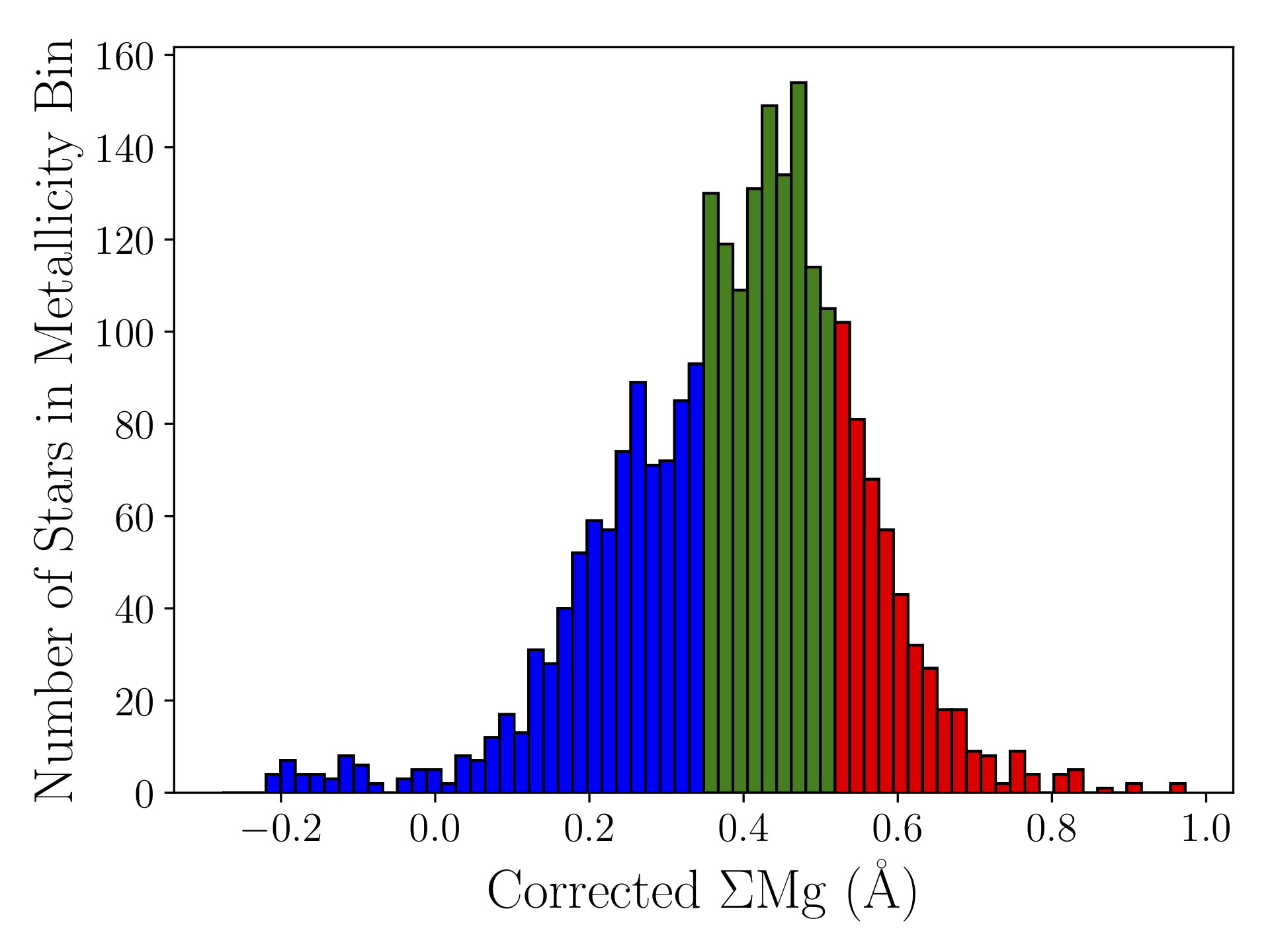}
     \caption{The corrected metallicities of stars in Fornax calculated using Equation \ref{eq:metCorr}.  The metal poor (MP), intermediate metallicity (IM), and metal rich (MR) subpopulations considered in this work are shown, respectively, in blue, green, and red.  The locations of the rigid metallicity cuts were based on the subpopulations of \cite{Amorisco13}.} \label{fig:metalCuts}
\end{figure} 

When we commenced this work, the data set of \cite{Walker09_data} contained the largest number of LG dSph stars with uniformly measured positions and velocities.  We used that data and the analyses described below to illustrate the capabilities of the method described in Section \ref{sec:theMethod}.  We hope that our analysis will help motivate further efforts to measure the kinematics of additional LG dSph stellar populations.
 
We used only those stars reported by \cite{Walker09_data} that have membership likelihoods of $\geq 0.75$, leaving us with 2498 RG stars.  As explained in \cite{Walker09_data}, we measured the corrected metallicity, $\Sigma Mg'$, from the raw measurement of the metallicity, $\Sigma Mg$, and from the apparent V-band magnitude, $V$, via a linear correction calibrated against the typical horizontal branch V-band magnitude, $V_{HB}$, of the galaxy of interest:
\begin{equation} \label{eq:metCorr}
\Sigma Mg ' = \Sigma Mg + (0.079 \pm 0.002)(V - V_{HB}) \ .
\end{equation}
For Fornax, we used the value $V_{HB} = 21.3$ reported in \cite{Walker11}. 

In our analysis of Fornax, we used several dSph properties not measured by \cite{Walker09_data}.  We list those properties in Table \ref{tab:fornExtras}. 

\begin{deluxetable} {C C C}
\caption{Used Fornax properties.}
\label{tab:fornExtras}
\tablehead{\colhead{Fornax dSph Property} &  \colhead{Value} & \colhead{Reference}}  
\startdata 
 \textrm{Distance from Milky Way, } D $ & $147\textrm{ kpc}$ &   (1) \\ 
 \textrm{Total stellar mass, }M_{\star} &   $10^{7.39} M_{\odot} $  &   (2) \\ 
 \textrm{Half-light radius, }r_{1/2} &  $791 \textrm{ pc}$ & (3)  \\
  \textrm{R.A.} &  $02\textrm{h:}39\textrm{m:}59\textrm{s} $ & (4)  \\
  \textrm{Decl.} &  $-34\textrm{d:}27\textrm{m:}00\textrm{s}$ & (4)  \\ 
  \textrm{R.A. proper motion, }\mu_{\alpha} \cos{(\delta)} &  376 \pm 3 \mu\textrm{as yr} ^ {-1}& (5)  \\ 
    \textrm{Decl. proper motion, }\mu_{\delta} &  -413 \pm 3 \mu\textrm{as yr} ^ {-1} & (5) 
 \enddata 
 \tablerefs{(1) \cite{McConnachie12}, (2) \cite{Kirby13}, (3) \cite{Munoz18}, (4) \cite{Walker11}, (5) \cite{Gaia18}}
\end{deluxetable}

LG dSphs contain distinct subpopulations of RG stars \citep{Kleyna03, McConnachie07, Walker11, Amorisco12, Agnello12, Lora13, Majewski13, Pace16, Caldwell17, Strigari17, Contenta17}.  Each subpopulation is characterized by distinct parameters and, provided it conforms to the appropriate assumptions, must satisfy Equation \ref{eq:genericJeans} independently.  Thus, each subpopulation can produce a distinct observed stellar profile while occupying the same underlying gravitational potential.  Subpopulation disambiguation was particularly valuable to our modeling efforts because stellar substructure that may be visible in isolated subpopulations can get washed out when subpopulations are lumped together.  The distinct morphologies of the different populations identified in \cite{Battaglia06, delPino15, Wang19}, which are not visible when all stars in Fornax are treated as a monolithic whole, underscore this point.

\begin{figure}
    \centering
    \includegraphics[width=0.5 \textwidth]{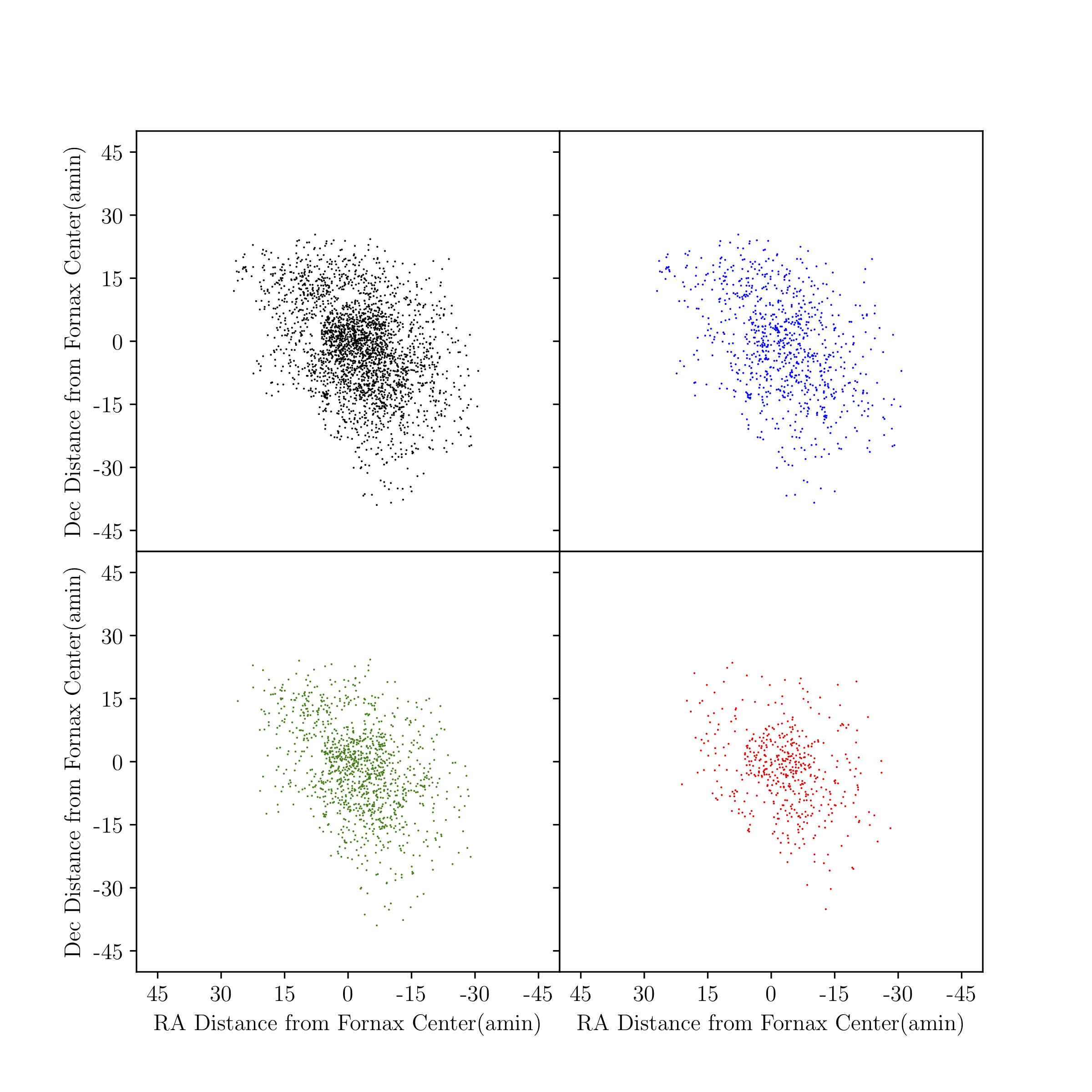}
    \caption{The red giant (RG) stars of Fornax, divided into our metal poor (MP), intermediate metallicity (IM), and metal rich (MR) subpopulations in the upper right, lower left, and lower right panels, respectively.  The upper left panel shows all used Fornax RG stars.  The subpopulations were determined using rigid metallicity cuts informed by the subpopulations identified by \cite{Amorisco13}.  Consistent with the expectations of Jeans' analysis, the MR subpopulation is the most centrally concentrated and is characterized by the smallest velocity dispersion.} \label{fig:fornStars}
\end{figure}

Because our method for measuring the correspondence between an observed data set and a surface brightness model (Section \ref{sec:compLikelihood}) requires that each star belong exclusively to one stellar subpopulation, and because we endeavor to use no presupposed stellar profiles, we used only rigid metallicity cuts to break the Fornax RG stars of \cite{Walker09_data} into distinct subpopulations.  Specifically, we defined a metal poor (MP) subpopulation, an intermediate metallicity (IM) subpopulation, and a metal rich (MR) subpopulation. 
We based the $\Sigma Mg'$ values of our metallicity cuts on the probabilistic metallicity membership functions of \cite{Amorisco13}.  We show the metallicity cuts in Figure \ref{fig:metalCuts} and the resulting subpopulations in Figure \ref{fig:fornStars}.  

\subsubsection{The Gravitational Potential of a Disk in a DM Halo} \label{sec:DDDM} 
Assuming that the relevant mass for determining the system's gravitational potential is described by mass density, $\rho (\mathbf{x})$, we determined the gravitational potential, $\Phi(\mathbf{x})$, by solving Poisson's equation:
\begin{equation}
    \nabla ^ 2\Phi(\mathbf{x})= -4 \pi G \rho (\mathbf{x}) \ ,
    \label{eq:poisson} 
\end{equation} 
where $G$ is Newton's Constant.  
The simulations of \cite{Battaglia15} suggest that a Fornax-like dSph would have its stellar population altered by only a few percent due to the tidal influence of the Milky Way, and we model Fornax as an isolated system. 

In this work, we separated the gravitationally dominant components of a dSph into two distinct elements: a spheroidal DM halo, $\rho_{h}(\mathbf{x})$, and an embedded disk, $\rho_{d}(\mathbf{x})$.  The disk could consist either of baryons or DM.  
The total mass distribution that determines the gravitational potential was thus
\begin{equation}
    \rho(\mathbf{x}) = \rho_{h}(\mathbf{x})  + \rho_{d}(\mathbf{x}) \ .
    \label{fullPot} 
 \end{equation}
 The linearity of Equation \ref{eq:poisson} allowed us to decompose the gravitational potential into halo and a disk components:
\beq
  \Phi(\mathbf{x}) = \Phi_h(\mathbf{x}) + \Phi_d(\mathbf{x}) \ ,
 \eeq
  where 
\begin{equation} 
   \nabla ^2 \Phi_h(\mathbf{x}) = - 4 \pi G \rho_{h}(\mathbf{x}) \ ,
\end{equation}    
and 
\begin{equation} 
\nabla^2 \Phi_d(\mathbf{x}) = - 4 \pi G\rho_{d}(\mathbf{x}) \ .
\end{equation} 

Although the findings of \cite{Fitts17} suggest that Fornax's DM halo may be triaxial, we assumed that it is spheroidal in partial accordance with the findings of \cite{Hayashi12, Klop17, Genina18, ElBadry18}.  We considered three DM halo models: a spheroidal version of the cusped DM halo that \cite{Navarro97} found to characterize the DM halos predicted by CDM simulations, a spheroidal version of the cored DM halo that \cite{Burkert95} found to predict the rotation curves of dwarf galaxies, and an alternate cored, spheroidal DM halo not associated with previous analyses.  We refer to these halos, respectively, as a spheroidal NFW halo, a spheroidal Burkert halo, and a spheroidal Acored halo.  For these three halos, 
\begin{equation} \label{eq:DMHDef}
  \rho_{h} = 
  \begin{cases} 
      \frac{\rho_s}{m'({1 + m'})^2} &  (\textrm{NFW}) \\
      \frac{\rho_s}{({1 + m'})({1 + m'} ^ 2)}  & (\textrm{\textrm{Burkert}})  \\
      \frac{\rho_s}{({1 + m'})^3}  & (\textrm{\textrm{Acored}}) \ ,
   \end{cases} 
\end{equation}
where $R$ is the distance off of the spheroid's axis of symmetry, $z$ is the distance along the polar axis, $r_s$ is the scale radius of the halo, $m'^2  \equiv (R/r_s)^2 + (z/(r_s Q))^2$, $\rho_s$ is the scale mass density of the halo, and $Q \in [0,\inf)$ is the fixed ratio of the polar and equatorial axes of the halo's equidensity shells.  We call $Q$ the `ellipticity' of the halo.  Note that $Q= 1$, $Q > 1$, and $Q < 1$ correspond respectively to spherical, prolate, and oblate halos.  

We defined the cutoff parameter, $c$, as the number of spheroidal scale radii from the dSph center where the $\rho_d$ discontinuously falls to $0$. 
We extended the models of Equation \ref{eq:DMHDef}, as they otherwise predict infinite DM halo mass.  Any extension that satisfies $\mathcal{O}(\rho_{DM}(m')) < \mathcal{O}(1/{m'}^3)$ would do, and we chose the simplest.

As we discuss in Appendix \ref{app:haloPot}, it was more convenient to parameterize $\rho_{h}$ by the total halo mass, $M_{h}$, than by $\rho_s$.  Integrating Equation \ref{eq:DMHDef} over the volume defined by $m' < c$ to determine $M_h$, we rewrote $\rho_{h}$ as 
\begin{equation} 
  \rho_{h} = 
  \begin{cases}
     \rho_{\textrm{int}} & m' \leq c \\
      0 & 0 > c \ ,
  \end{cases}
  \label{eq:HaloDens}
\end{equation} \label{eq:DMHUsedInt}
where 
\begin{equation} 
  \rho_{\textrm{int}} = \frac{M_h}{4 \pi Q r_s ^ 3}
  \begin{cases} 
       \frac{1}{ f(c)(m' (1 + m') ^2)} &  (\textrm{NFW}) \\
     \frac{1}{ h(c)(1 + m') ^3}   & (\textrm{\textrm{Acored}}) \\
      \frac{1}{g(c)(1 + m') (1 + m'^2)}   & (\textrm{\textrm{Burkert}}) \ ,
   \end{cases} 
   \label{eq:HaloInt}
\end{equation} 
and where $f(c) \equiv \ln (1+c) - c/(1+c)$, $h(c) \equiv \ln(1+c) - (2c + 3 c^2)/(2(1+c)^2)$, and $g(c) \equiv 1/2 ( \ln{(1+c)} + \arctan{(c)})$ are scaling constants related to the cutoff radius of the halo (see Appendix \ref{app:haloPot} for details).  

There are various methods for computing $c$.  The standard method for spherical halos, and the one that we repurposed for spheroidal halos, is to set $cr_s$ equal to the cosmological virial radius \citep{Kazantzidis06}.  Specifically, we set $cr_s$ equal to the spheroidal radius within which the average DM density, $M_h/(4\pi/3\ Q(c\ r_s) ^ 3)$, is equal to the present day critical mass density of the universe, $\rho_{crit,0} = H_0^2 /(8\pi /3 \ G) $, multiplied by a scalar overdensity constant, $\Delta_c$: 
\begin{equation} \label{eq:cutoffRad} 
\frac{M_h}{4\pi/3\ Q (c\ r_s) ^ 3} = \Delta_c \rho_{crit,0} \Rightarrow c ^ 3 = \frac{2 G M_h}{Q \Delta_c H_0 ^ 2 r_s ^ 3 } \ ,
\end{equation}   
where $H_0$ is the Hubble Constant.  We used the value of $H_0 = 67.31 \pm 0.96$ reported by the \cite{Planck16}, and the standard value of $\Delta_c = 200$ discussed by \cite{Navarro96Symposium, White00}.  

For every choice of $M_h$ and $r_s$, we used Equations \ref{eq:HaloDens}, \ref{eq:HaloInt}, and \ref{eq:cutoffRad} to determine the DM halo mass density, $\rho_h$.  

We modeled $\rho_{d}$ as a hyperbolic secant disk, similar to that used by \cite{Robin14} to parameterize the disk of the Milky Way:
\begin{equation}
    \rho_{d} (R,z)= \frac{M_d}{8 \pi R_d^2 z_d} \exp(-\frac{R}{R_d}) \sech(\frac {z}{2 z_d}) \ ,
    \label{eq:DiskDens}
\end{equation}
where $R$ is the distance off of the polar axis, $z$ is the distance along the polar axis, $M_d$ is the disk mass, and $R_d$ and $z_d$ respectively the disk's scale radius and scale height. 

Inserting Equations \ref{eq:HaloDens} and \ref{eq:DiskDens} into Equation \ref{eq:poisson}, we acquired a pair of differential equations to solve.  Via the methods outlined in Appendices  \ref{app:haloPot} and \ref{app:diskPot}, we expressed the solutions to Equation \ref{eq:poisson} for the halo and disk as 
\beq \label{eq:fullHaloPot} 
\Phi_{h}(u, v, M_h, r_s, Q, c) 
=   \frac{- G M_h}{r_s} \ \Gamma (u, v, Q, c) 
\eeq
and 
\beq     \label{diskPot}
    \Phi_{d} (\alpha, \beta, M_d, \lambda, z_d, c) = -\frac{G M_d}{z_d} F(\alpha,\beta,\lambda, c) \ , 
\eeq
where $\alpha \equiv R/R_d$, $\beta \equiv z/z_d$, $\lambda \equiv z_d/R_d$, $u \equiv R/r_s$, $v \equiv z/r_s $, and $\Gamma$ and $F$ are single integral functions defined respectively in Equations \ref{eq:fullHaloPot} and \ref{eq:fullDiskPot}.  $\Gamma$ and $F$ were computed numerically.  

We placed these gravitational potential profiles in the coordinates of an earth-based observer, $\mathbf{x}_{\textrm{sky}} = (x_{\textrm{sky}},y_{\textrm{sky}},z_{\textrm{sky}})$, where $x_{\textrm{sky}}$ is distance east on the sky, $y_{\textrm{sky}}$ is distance directly away from the observer, $z_{\textrm{sky}}$ is distance north on the sky, and where $\mathbf{x}_{\textrm{sky}}= (0,0,0)$ at the center of Fornax listed in Table \ref{tab:fornExtras}.  
Allowing the disk and halo to sit at independent and arbitrary orientations, we defined the two sets of spherical angles $(\psi,\theta)$ and $(b, a)$ to describe, respectively, the orientation of the polar axes of the halo and of the disk in the observer's coordinate system.  We further allowed independent offsets, $\mathbf{h_c}$ and $\mathbf{d_c}$, between the reported dSph center and the centers of the halo and disk.  The natural coordinates of the halo, $\mathbf{x_h} = (x_h, y_h, z_h)$, are related to the observer's coordinates by
\beq\label{eq:hToSkyCoords}
    \mathbf{x_h}  &(\mathbf{x_{\textrm{sky}}} , \theta, \psi,\mathbf{h_c} )  = \mathbf{h_c} \\ 
    & + [ \cos{(\theta)} (x_{\textrm{sky}} \cos{(\psi)} + y_{\textrm{sky}} \sin{(\psi)}) - z_{\textrm{sky}} \sin{(\theta)},\\
                                                                                        &\ \ \ \   - x_{\textrm{sky}} \sin{(\psi)} + y_{\textrm{sky}} \cos{(\psi)}, \\
                                                                                        &\ \ \ \ \sin{(\theta)} (x_{\textrm{sky}} \cos{(\psi)} + y_{\textrm{sky}} \sin{(\psi)})+ z_{\textrm{sky}}\cos{(\theta)} ] ,
\eeq
and the natural coordinates of the disk, $\mathbf{x_d}=(x_d, y_d, z_d)$, are related to the observer's coordinates by
\beq\label{eq:dToSkyCoords}
    \mathbf{x_d}  &(\mathbf{x_{\textrm{sky}}} , \theta, \psi,\mathbf{d_c} ) =  \mathbf{d_c} \\ 
    & + [ \cos{(a)} (x_{\textrm{sky}} \cos{(b)} + y_{\textrm{sky}} \sin{(b)}) - z_{\textrm{sky}} \sin{(a)},\\
                                                                                        &\ \ \ \   - x_{\textrm{sky}} \sin{(b)} + y_{\textrm{sky}} \cos{(b)}\ , \\
                                                                                        &\ \ \ \ \sin{(a)} (x_{\textrm{sky}} \cos{(a)} + y_{\textrm{sky}} \sin{(a)})+ z_{\textrm{sky}}\cos{(b)} ] \ . 
\eeq
The unitless coordinates of the halo are 
\beq \label{eq:cylDiskToSky}
        u(& \mathbf{x_{\textrm{sky}}}, r_s, \psi, \theta, \mathbf{h}_{c})  \\ 
        & = \frac{(x_h^2(\mathbf{x}_{\textrm{sky}}, \theta, \psi,\mathbf{h}_{c} ) + y_h^2(\mathbf{x}_{\textrm{sky}}, \theta, \psi,\mathbf{h}_{c} ) )^{1/2}}{r_s}  ,\\
        v(& \mathbf{x}_{\textrm{sky}}, r_s, \psi, \theta, \mathbf{h}_{c}) = \frac{z_h(\mathbf{x}_{\textrm{sky}}, \theta, \psi,\mathbf{h}_{c} ) }{r_s} \ ,
\eeq
and the unitless coordinates of the disk are 
\beq \label{eq:cylHaloToSky}
        \alpha(\mathbf{x}_{\textrm{sky}}, & r_s, b,a, \mathbf{d}_{c},\lambda) \\ 
        & = \frac{1}{\lambda r_s} (x_d^2(\mathbf{x}_{\textrm{sky}}, b, a,\mathbf{d}_{c} ) + y_d^2(\mathbf{x}_{\textrm{sky}}, b, a,\mathbf{d}_{c} ) )^ {1/2} \ ,\\
        \beta(\mathbf{x}_{\textrm{sky}}, & r_s, b,a, \mathbf{d}_{c},\lambda, R_d) \\ 
        & = \frac{1}{R_d \lambda } z_d(\mathbf{x}_{\textrm{sky}}, b, a,\mathbf{d}_{c} ) \ . 
 \eeq

Combining these various coordinate parameters, we found a final expression for the total gravitational potential of a DM distribution consisting of a spheroidal halo and a disk: 
\beq \label{eq:fullPotFunct}
\Phi(& \mathbf{x}_{\textrm{sky}}, M, r_s, Q, \psi, \theta, R_d, \epsilon, \lambda, b, a, \mathbf{h}_{c}, \mathbf{d} _{c}, c) \\
 & = \Phi_{h}   (u(\mathbf{x}_{\textrm{sky}}, r_s, \psi, \theta, \mathbf{h}_{c}), \\
  &\ \ \ \ \ \ \ \ \ v(\mathbf{x}_{\textrm{sky}},r_s, \psi, \theta, \mathbf{h}_{c}),(1-\epsilon) M, r_s, Q, c) \\ 
&+ \Phi_{d}(\alpha(\mathbf{x}_{\textrm{sky}}, r_s, b, a, \mathbf{d}_{c}, R_d), \\ 
&\ \ \ \ \ \ \ \ \beta(\mathbf{x}_{\textrm{sky}}, r_s, b, a, \mathbf{d}_{c}, R_d, \lambda), \epsilon M, \lambda, R_d \lambda, c  ) \\
 & = -\frac{G M}{r_s}  \Big [(1 - \epsilon) \Gamma(u(\mathbf{x}_{\textrm{sky}},r_s, \psi, \theta, \mathbf{h}_{c}), \\ 
 & \ \ \ \ \ \ \ \ \ \ \ \ \ \ \ \ \ \ \ \ \ \ \ \ \ \ v(\mathbf{x}_{\textrm{sky}},r_s, \psi, \theta, \mathbf{h}_{c}), Q, c)  \\
& \ \ \ \ \ \ \ \ \ \ \ \ \ + \frac{r_s \epsilon }{R_d \lambda} F(\alpha(\mathbf{x}_{\textrm{sky}}, r_s, b,a, \mathbf{d}_{c},\lambda), \\ 
& \ \ \ \ \ \ \ \ \ \ \ \ \ \ \ \ \ \ \ \ \ \ \ \  \beta(\mathbf{x}_{\textrm{sky}}, r_s, b,a, \mathbf{d}_{c}, \lambda), \lambda, c) \Big ] \ , 
\eeq
where $\epsilon$ is the fraction of the matter in the disk: 
\beq
\epsilon \equiv M_d / M \Rightarrow M_h = (1- \epsilon) M \ . 
\eeq

We used the parameter vector, $\mathbf{\Theta}$, to hold all parameters other than the sky coordinates:   
\beq
\mathbf{\Theta} = \{M, r_s, Q, \psi, \theta, R_d, \epsilon, \lambda, b, a, \mathbf{h}_{c}, \mathbf{d} _{c}, c \} \ .
\eeq
We varied differing subsets of the parameters in $\mathbf{\Theta}$ as part of the MCMC analysis of Section \ref{sec:MCMC}. 

We note that the independently oriented disk and halo in our expression violate the assumption of spheroidal symmetry used in the derivation of Equation \ref{eq:JEqSol}.  For the sake of computational convenience, we still used that analytical result, but we recognize that a more accurate examination of this combined disk and halo model would require one to insert this truly three-dimensional gravitational potential into the original Jeans equations.  Note that this violation of symmetry applies only to the analysis that permits the existence of a disk.  

\subsubsection{Measuring Kinematical Parameters of Fornax}\label{sec:fornKinematics}

\begin{deluxetable} {C C}
\caption{The Directly Measured Values of $\sigma ^2$ for Each Population }
\label{tab:velDisp}
\tablehead{\colhead{Population} &   $\sigma^2 $ ($\textrm{km} ^ {2} \textrm{s} ^ {-2}$) \textrm{ (corrected)}  } 
\startdata 
 \textrm{\ \ \ \ \ \ \ \ \ \ \ \ \ \  MR\ \ \ \ \ \ \ \ \ \ \ \ \ \   } & $\ \ \ \ \ \ \ \ \ \ \ \ \ \   102.2 \pm{2.1}\ \ \ \ \ \ \ \ \ \ \ \ \ \    \\ 
 \textrm{\ \ \ \ \ \ \ \ \ \ \ \ \ \   IM\ \ \ \ \ \ \ \ \ \ \ \ \ \   } &  $\ \ \ \ \ \ \ \ \ \ \ \ \ \    130.4 \pm{1.4}\ \ \ \ \ \ \ \ \ \ \ \ \ \     \\ 
 \textrm{\ \ \ \ \ \ \ \ \ \ \ \ \ \   MP\ \ \ \ \ \ \ \ \ \ \ \ \ \    } &  $\ \ \ \ \ \ \ \ \ \ \ \ \ \  156.2 \pm{2.9}\ \ \ \ \ \ \ \ \ \ \ \ \ \  
 \enddata 
 \tablecomments{For the three considered red giant populations in Fornax listed in Column 1, Column 2 contains the velocity dispersion, $\sigma ^2$, after correcting for the dSph's proper motion.  
The uncertainties were determined by propagating the reported errors in the heliocentric velocities and in the dSph's bulk motion on the sky through Equation \ref{eq:sigSqrEq}. 
The average velocity dispersion measurements reported by \cite{Amorisco13} for the MR, IM, and MP populations are, respectively, smaller than, consistent with, and larger than the velocity dispersions measured here.  Because \cite{Amorisco13} used the velocity of the stars to  determine their populations, their stellar populations are more kinematically distinct.} 
\end{deluxetable}

For a given potential, we modelled Fornax using the solution to the Jeans equations discussed in Appendix \ref{app:axisymPot}.  Within this analysis paradigm, the kinematics of stellar subpopulation $A$ are fully characterized by two parameters: the constant velocity dispersion of the population, $\sigma_A$, and the constant angular velocity of the population, $\omega_A$.  For Fornax, we had only the line of sight velocities, $v_r$, of individual member stars from which to infer these quantities.  

If a dSph (or any extensive astronomical object) has proper motion, $\mathbf{w}$, the projection of this proper motion onto the observer's line of sight is position dependent.  Defining $\mathbf{v'_n} $ as the velocity of star $n$ with respect to the dSph as a whole (the value of interest for calculating $\sigma^2_a$ and $\omega_A$), the radial velocity of that star with the dSph proper motion effect removed is
\beq \label{eq:defCorrV} 
v'_{r,n} &=\mathbf{v'_n} \cdot \hat{r} = (\mathbf{v_n} - \mathbf{w}) \cdot \hat{r}  \\ 
&\simeq v_{r,n} - \frac{w_x x_n + w_y D + w_z z_n}{\sqrt{x_n^2 + D^2 + z_n^2}} \\ &
\simeq v_{r,n} - (w_x \frac{x_n}{D} + w_y + w_z \frac{z_n}{D} ) \ , 
\eeq
where $v_{r,n}$ is the measured radial velocity of star $n$, $x_n$ and $z_n$ are the physical distances from star $n$ to the center of the galaxy projected onto the sky, $D$ is the distance from the observer to the dSph's center, and $x_n^2 + D^2 +z_n^2 \simeq D^2$ since the distance from an earth-based observer to a dSph far exceeds the size of the dSph.

The transverse velocity of the dSph is related to its measured angular velocity by
\beq
 (w_x, w_z) = (\mu_{\alpha} \cos{(\delta)} D, \mu_{\delta} D) \ ,
\eeq
where the tangential angular velocity of the dSph, $(\mu_{\alpha} \cos{(\delta)}, \mu_{\delta})$, are listed in Table \ref{tab:fornExtras} and must be recast in radians per unit time.  
Note that $w_y$, as an overall additive term, did not affect our measurements of $\sigma_A^2$ and $\omega_A$. 

For population $A$, we measured $\sigma_A^2$ from the corrected radial velocities, $v_{r,n}'$ in the usual way:
\begin{equation} \label{eq:sigSqrEq}
\sigma_A^2 = \frac{1}{N_{A} }\sum_{ n \in A} (<v_{r}'>_{A} - v_{r,n}') ^ 2 \ ,
\end{equation}
where the sum and average are taken over all stars in subpopulation $A$.  
The resulting values of $\sigma_A^2$ for the three Fornax populations are listed in Table \ref{tab:velDisp}.  

\begin{figure}
    \centering
    \includegraphics[width=0.5 \textwidth]{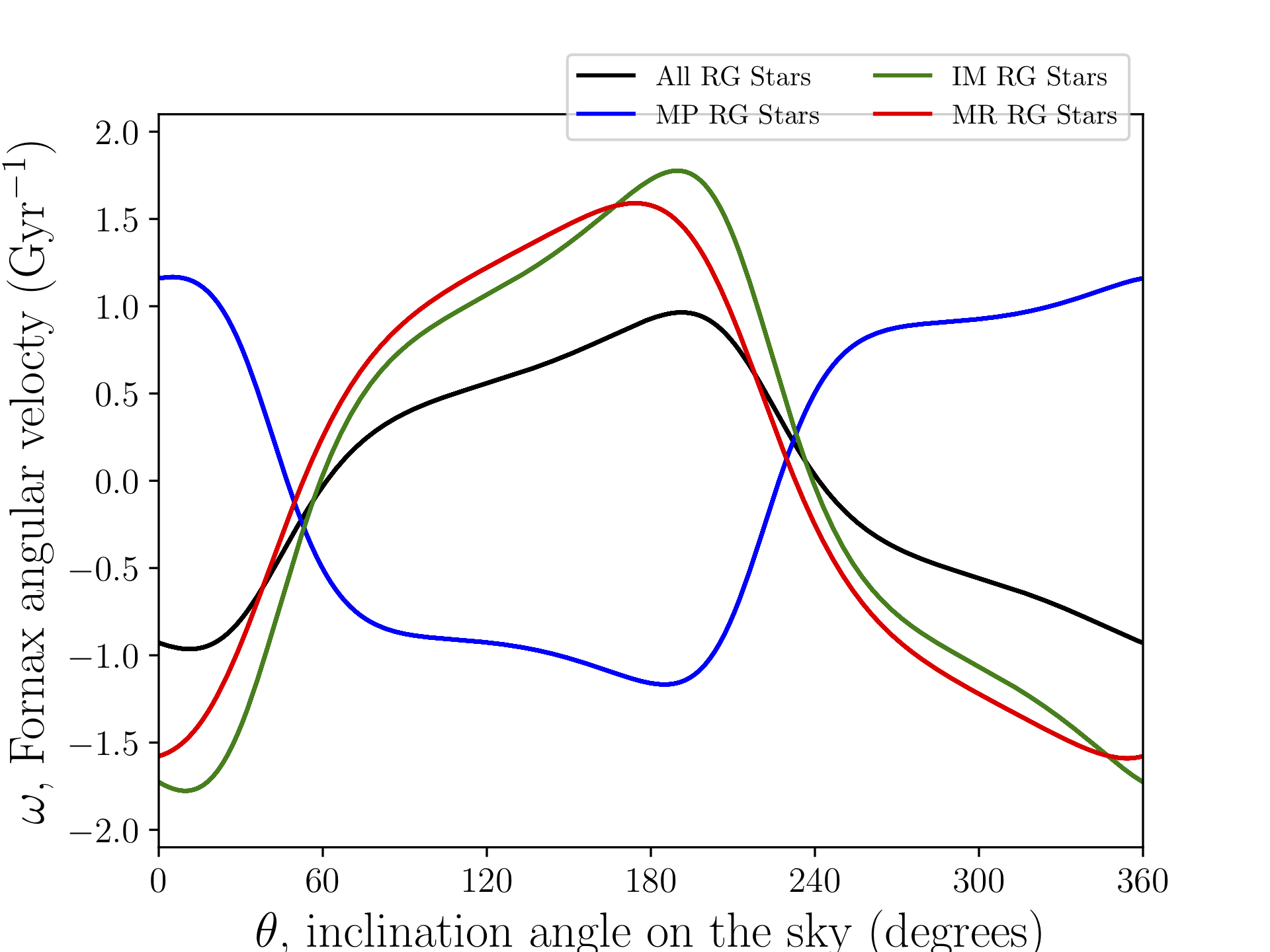}
    \caption{The rotation signal for each of Fornax's RG populations as a function of rotation direction on the sky, under the assumption that Fornax's center of mass matches the photometric center reported by \cite{Walker11} ($\mathbf{h_c} = 0$) and that the polar-axis lies in the plane of the sky ($\psi = 0$).  The geometry of a spheroidal distribution of stars can produce an artificial rotation signal when the direction of rotation is improperly chosen.  Thus, the largest rotation signal, which occurs roughly at $\theta = 180^{\circ}$, should not be interpreted as the true rotation signal.  \label{fig:rotThroughCenter}}
\end{figure}

We measured the bulk rotation of population $A$ in Fornax by fitting a line to the corrected line of sight velocities as a function of position perpendicular to the polar axis of the DM halo:
\beq
& (\omega_X, w_{y, X}) \\ 
&= \underset{p_1 \in R, p_0 \in R} {\textrm{argmin}} \Big (\sum_{ n \in X} \frac{1}{\sigma_{v_{r,n}}^2} (v_{r,n}' - (p_1 R_n + p_0))^2 \Big ) \ ,
\eeq
where $p_0$ and $p_1$ are fitting parameters of the minimization and where $R_n = u_n r_s$ is the distance of the $n^{\textrm{th}}$ star in subpopulation $A$ from the symmetry axis of the dSph, calculated using Equations \ref{eq:hToSkyCoords} and \ref{eq:cylDiskToSky}. 
Because $\omega_A$ depends on the orientation and center of the dSph, it needed to be recomputed for all values of $\psi$ and $\mathbf{h_c}$, as defined in Section \ref{sec:DDDM}.  In Figure \ref{fig:rotThroughCenter}, we show $\omega$ for the Fornax stellar populations at various orientation angles on the sky when the galaxy's center of light is assumed equal to its center of mass ($\mathbf{h_c} = 0$) and when the polar axis is assumed to lie in the plane of the sky ($\psi = 0$).  

If the direction of rotation is misidentified, the geometry of the dSph may artificially increase or decrease the measured value of $\omega$.  As with the other model parameters, an incorrect rotation signal produced an inferior fit to the observed data.  One should not assume that the orientation with the largest rotation signal is necessarily the orientation of the true rotation.  

\subsubsection{Measuring the On Sky Probability Distribution} \label{sec:losApp} 
Using the gravitational potential of Section \ref{sec:DDDM}, the kinematical parameters measured in Section \ref{sec:fornKinematics}, and the Jeans Equations solution of Equation \ref{eq:JEqSol}, we calculated the volume probability density of Fornax, $\nu$, for a posited DM density in the three-dimensional coordinates of the observer, $\mathbf{x}_{\textrm{sky}}$.  However, we were interested in the two-dimensional projection of $\nu$ onto the sky of an observer on Earth: 
\beq
S(x_{\textrm{sky}},& z_{\textrm{sky}},\mathbf{\Theta}) \\ 
&= \int_{-D}^{\infty} \nu(x_{\textrm{sky}},y_{\textrm{sky}},z_{\textrm{sky}},\mathbf{\Theta}) d y_{\textrm{sky}} \ . 
\eeq
Recall that $x_{\textrm{sky}}$ and $z_{\textrm{sky}}$ are the on-sky coordinates centered on the galaxy with $x_{\textrm{sky}}$ running parallel to astronomical right ascension (RA) and $z_{\textrm{sky}}$ running parallel to astronomical declination (Dec), that $y_{\textrm{sky}}$ is the line of sight distance from the Earth to a given spatial slice of the integrand, and that $D$ is the distance from Earth to the center of the dSph.  

Because the gravitational potential, $\Phi$, was computed numerically, we calculated $\nu$ at a finite set of points.  We approximated the integral using a Riemann sum:
\beq \label{eq:losInt}
S&(x_{\textrm{sky}},z_{\textrm{sky}},\mathbf{\Theta}) 
   = \int_{-D}^{\inf}  \nu(x_{\textrm{sky}},y_{\textrm{sky}},z_{\textrm{sky}},\mathbf{\Theta}) d y_{\textrm{sky}} \\
&\simeq \sum_{k = 0}^{N_{\textrm{los}}-1} \nu(x_{\textrm{sky}},\frac{1}{2}(y_{\textrm{sky},k} + y_{\textrm{sky},k+1}),z_{\textrm{sky}},\mathbf{\Theta}) \\ 
    &\ \ \ \ \ \ \ \ \ \  \ \times (y_{\textrm{sky},k+1} - y_{\textrm{sky},k}) \\ 
    & = \tilde S(x_{\textrm{sky}}, z_{\textrm{sky}}, \mathbf{\Theta}, N_{\textrm{los}}) , 
\eeq
where $N_{\textrm{los}}$ is the number of planes along the line of sight on which $\nu$ was explicitly calculated.  
For a Riemann sum, 
\beq
& | S (x_{\textrm{sky}},z_{\textrm{sky}},\mathbf{\Theta}) - \tilde S(x_{\textrm{sky}}, z_{\textrm{sky}}, \mathbf{\Theta}, N_{\textrm{los}}) | \\   \ &\ \ \ \ \  \leq \textrm{max} \Big (\frac{d^2 \nu(x_{\textrm{sky}},y_{\textrm{sky}},z_{\textrm{sky}},\mathbf{\Theta})  }{d_{\textrm{sky}}^2} \Big ) \\ & \ \ \ \ \ \ \ \ \ \ \ \ \ \ \ \ \  \times \frac { ( y_{\textrm{sky},N_{\textrm{los}}-1} - y_{\textrm{sky},0} ) ^ 3}{24 N_{\textrm{los}}^2}  ,
\eeq
provided that there is negligible mass density beyond the line of sight bounds, $(y_{\textrm{sky},0}, y_{\textrm{sky},N_{\textrm{los}}-1})$.

We were restricted to computing the discrete sum in Equation \ref{eq:losInt} at only a finite set of $(x_{\textrm{sky}},z_{\textrm{sky}})$ points.  We set up a grid of sky coordinates, $(x_{\textrm{sky},i},z_{\textrm{sky},j})$, where $i$ and $j$ are integers in the ranges $[0,N_x - 1],[0,N_z - 1]$, calculate $\tilde S_{i,j}(\mathbf{\Theta}) \equiv \tilde S(x_{\textrm{sky},i},z_{\textrm{sky},j},\mathbf{\Theta})$ for each point on the grid, and approximated $S(x_{\textrm{sky}},z_{\textrm{sky}},\mathbf{\Theta})$ by interpolating linearly over the $\tilde S_{i,j}(\mathbf{\Theta})$ grid.  
We chose $(x_{\textrm{sky},0}, x_{\textrm{sky},N_x-1})$ and $(z_{\textrm{sky},0}, z_{\textrm{sky},N_z-1})$ to fully cover the observed region of the sky, and set $y_{\textrm{sky},0} = -50 \textrm{ arcmin}$ and $y_{\textrm{sky},N_y-1} = 50 \textrm{ arcmin}$.  We set the density of sampled points along each axis equal to 2 arcminutes: $(x_{\textrm{sky},N_x-1} - x_{\textrm{sky},0})/N_x = (y_{\textrm{sky}, N_y-1} - y_{\textrm{sky},0})/N_y = (z_{\textrm{sky},N_z-1} - z_{\textrm{sky},0})/ N_z = 2 \textrm{ arcmin}$ . 

\begin{figure} 
    \centering
        \includegraphics[width=0.5 \textwidth]{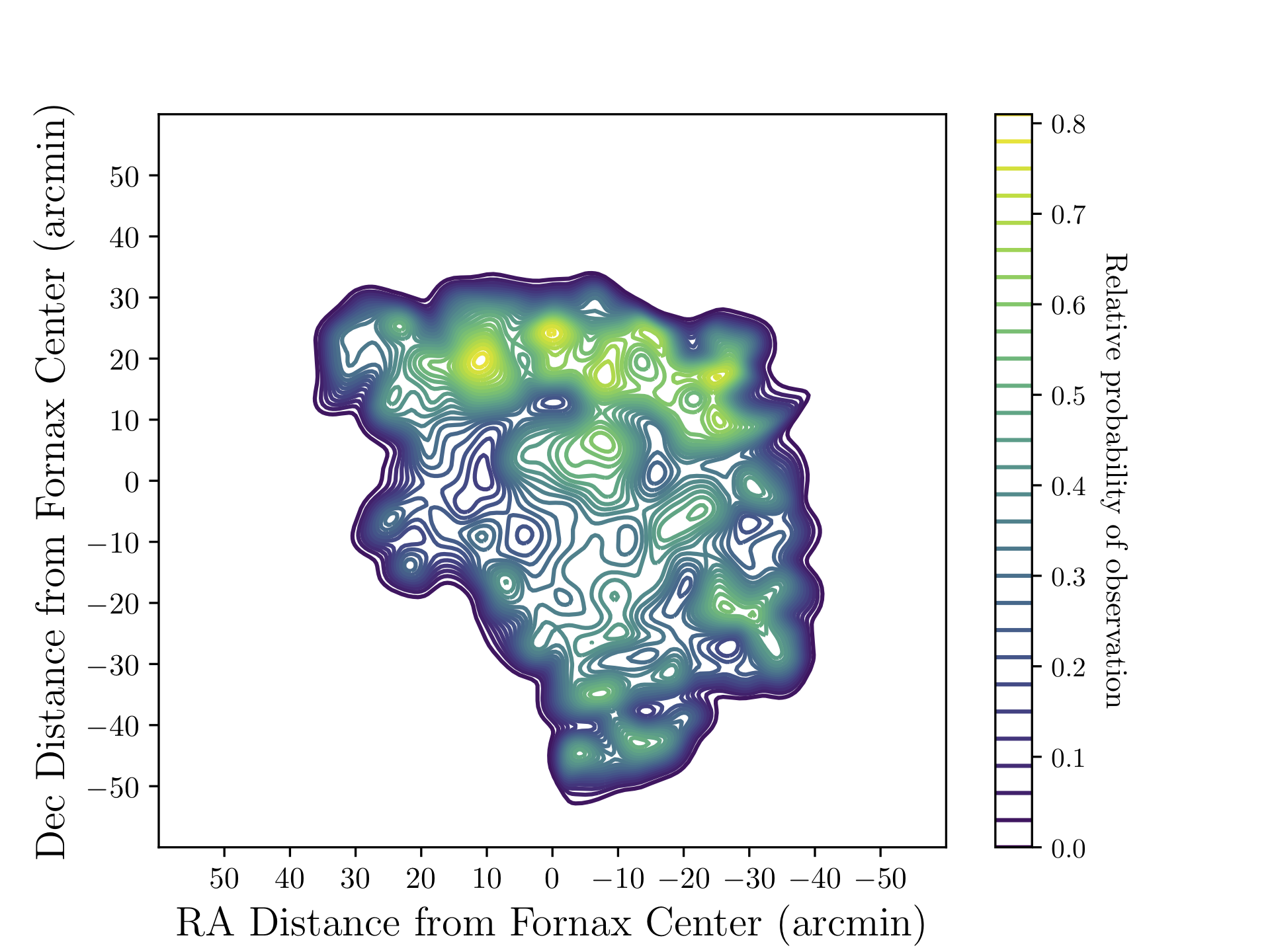} 
        \caption{The detectability function,  $D(x_{\textrm{sky}},  z_{\textrm{sky}})$, for the \cite{Walker09_main} observations of Fornax.  
        Comparing this detectability function to the observation fields in Figure 1 of \cite{Walker09_main} (we believe field 10 of that Figure is mislabeled), we see that regions of the sky characterized by overlapping fields have high values of $D(x_{\textrm{sky}},  z_{\textrm{sky}})$, while the unobserved regions between fields have low values.}  \label{fig:mask}
\end{figure}

We accounted for the uneven measurement of candidate Fornax RG stars by setting $D(x_{\textrm{sky}},z_{\textrm{sky}})$ equal to the position-dependent fraction of candidates for which good line of sight velocities were successfully measured.  Modifying the method discussed in Section 2.3 of \cite{Walker11}, we approximated this fraction by applying a Gaussian kernel to the lists of candidate and successfully measured stars:
\beq
& D(x_{\textrm{sky}},z_{\textrm{sky}})  = \frac{dN_{\textrm{obs}}(x_{\textrm{sky}},z_{\textrm{sky}})}{dN_{\textrm{cand}}(x_{\textrm{sky}},z_{\textrm{sky}})} \\
& \approx 
\frac{\displaystyle \sum_{i=1}^{N_{\textrm{obs}}} \exp  \Big [ -\frac{1}{2} \frac{(x_i - x_{\textrm{sky}})^2 + (z_i - z_{\textrm{sky}})^2 } {k_1^2} \Big ]}{\displaystyle \sum_{i=1}^{N_{\textrm{cand}}} \exp \Big  [ -\frac{1}{2} \frac{(x_i - x_{\textrm{sky}})^2 + (z_i - z_{\textrm{sky}})^2 } {k_1^2} \Big]} \ ,
\eeq
where $dN_{\textrm{obs}}(x_{\textrm{sky}},z_{\textrm{sky}})$ and $N_{\textrm{obs}}$ are respectively the number density and total number of observed stars with good velocity measurements, where $dN_\textrm{{cand}}(x_{\textrm{sky}},z_{\textrm{sky}})$ and $N_{\textrm{cand}}$ are respectively the number density and total number of candidate stars originally considered by \cite{Walker09_data}, and where $k_1$ is the kernel size of our smoothing.  Again following the example of \cite{Walker11}, we chose $k_1 = 2\textrm{ arcmin}$.  
We show the contours of $D(x_{\textrm{sky}},z_{\textrm{sky}})$ in Figure \ref{fig:mask}.  



We considered including the contribution of field-crowding in our calculation of $D(x_{\textrm{sky}},z_{\textrm{sky}})$ via the method discussed in the Appendix of \cite{Irwin84}.  However, we determined that such a correction would change $D(x_{\textrm{sky}}, z_{\textrm{sky}})$ by no more than one percent and would add considerably to our computation time.  We thus forwent this correction in our final analyses. 

We calculated, $O(x_{\textrm{sky}},z_{\textrm{sky}},\mathbf{\Theta})$, the probability of a star being observed at position $(x_{\textrm{sky}},z_{\textrm{sky}})$ for parameter set $\mathbf{\Theta}$, from $D(x_{\textrm{sky}},z_{\textrm{sky}})$ and $S(x_{\textrm{sky}},z_{\textrm{sky}}, \mathbf{\Theta})$ using Equation \ref{eq:defObserveProb}.

\subsubsection{Exploring the Parameter Space} \label{sec:MCMC}
We used a standard Markov Chain Monte Carlo (MCMC) sampling method to determine the best-fit parameters for a given DM model.  

For subsets of the model parameters in $\mathbf{\Theta}$, we ran a series of weighted random walks (`chains') through some allowed parameter range.  At the beginning of each step, a new point in the parameter space was chosen according to some sampling algorithm defined for each free parameter, centered at the current value.  A random number between $0$ and $1$ was drawn from a flat distribution.  If the likelihood ratio of the new point and the old point, as determined by Equation \ref{eq:relLike}, is greater than the random number, then the algorithm moved to the new point.  Otherwise, the step ended at the same point at which it began.  This method repeated a fixed number of times (typically $20000$).  The termination of every step was tracked over the course of the chain.  To help ensure that our algorithm had a chance to cover the complete parameter space, every hundredth step checked a distance ten times larger than normal when looking for a new position to which to move.  

Once several chains with various starting points (`seeds') were run, they were combined together to define the full set of raw results.  Such results constitute a series of $\sim 1.5 \times 10^6$ steps in the considered parameter space.  A parameter space may possess multiple local extremum to which the chains can converge.  Each of these points of convergence can correspond to a region of physical importance (such as the best fit prolate and oblate halos), and we thus did not force all chains to converge to the global extremum.  

For Fornax, we applied this MCMC to six different scenarios: a NFW DM halo without a disk (NWOD), a Burkert DM halo without a disk (BWOD), an Acored DM halo without a disk (AWOD), a NFW DM halo with a disk (NWD), a Burkert DM halo with a disk (BWD), and an Acored DM halo with a disk (AWD).  The scenarios without a disk were characterized by the parameter set $\mathbf {\Theta_{\textrm{WOD}}} = (M, r_s, Q, \psi, \theta, h_{c,x}, h_{c,y}, h_{c,z} )$, and the scenarios with a disk were characterized by the parameter set $\mathbf {\Theta_{\textrm{WD}}} = (M, r_s, Q, \psi, \theta, h_{c,x}, h_{c,y}, h_{c,z}, R_d, \lambda, b, a, d_{c,x}, d_{c,y}, d_{c,z})$.  We list the parameter ranges and step sizes used in the MCMC analysis in Tables \ref{tab:WODMCMCChain} and \ref{tab:WDMCMCChain}.

We did not allow the DM distribution centers to shift along the viewing axis ($h_{c,y} = d_{c,y} = 0$), as such an effect would have been wholly degenerate with changes in other parameters.  We also fixed additional parameters for computational reasons. 

For the variables $\{M, r_s, h_{c,x}, h_{c,z}, R_d, \lambda, d_{c,x}, d_{c,z}\}$, the new value for each step was determined by pulling a random number from a Gaussian distribution of a fixed width specific to each variable, centered at the old parameter value.  These values were constrained to some fixed bounds specific to each variable.  

For the ellipticity of the halo, $Q$, the sampling was complicated by the fact that the values of $Q \in [1, \inf)$ and $1/Q \in [1,\inf)$ define the same number of physical configurations (prolate vs oblate halos of varying ellipticities), but cover different sizes of the $Q$ space.  Therefore, we defined a surrogate value, $q$, that is related continuously to $Q$ by
\begin{equation} \label{eq:surrogateQ}
q(Q) = 
\begin{cases}
    Q & Q \geq 1 \\
    2-\frac{1}{Q} & Q < 1
\end{cases} \ , 
\end{equation}
from which it followed that 
\begin{equation}
\lim_{Q \rightarrow 1^+} q(Q) = \lim_{Q \rightarrow 1^-} q(Q) = 1 \ ,
\end{equation} 
that 
\begin{equation}
\lim_{Q \rightarrow \infty} q(Q) - q(1) = - (\lim_{Q \rightarrow 0} q(Q) - q(1)) \ ,
\end{equation} 
and that 
\begin{equation}
\begin{cases}
    \frac{dq}{dQ} = 1 & Q \geq 1 \\
    \frac{dq}{d(1/Q)} = -1 & Q < 1 \ .
\end{cases} 
\end{equation} 
Thus, Equation \ref{eq:surrogateQ} continuously maps $Q \in (0,1]$ and $Q \in [1,\inf)$ to ranges of $q$ with identical sizes and densities.  We sampled the $q$ parameter as we did the other scalar parameters, and converted $q$ back to $Q$ when describing the best fit distributions.  

The density of halo and disk orientations are not equally distributed over the space of orientation angles, $\{\psi,\theta, b,a\}$.  Determining the new halo orientation in each step by selecting new angles directly from a Gaussian distribution would have biased our results.  We instead defined each orientation by a unit 3-vector, $\hat{o}$.  With each step in the MCMC, we randomly perturbed each of the components of $\hat{o}$ by drawing from three flat distributions of equal width centered at the locations of the old coordinates, renormalized the perturbed $\hat{o}$, and took the new $\hat{o}$ as the orientation vector for the new point in the chain.  Because the disk and halo models considered have spheroidal symmetry, $\hat{o}$ and $-\hat{o}$ describe the same configuration.  We removed this degeneracy by forcing $\hat{o}_x \geq 0$.  Our sampling algorithm mapped any $\hat{o}$ vectors with $\hat{o}_x < 0$ to the identical configuration described by the vector $-\hat{o}$.  The angles $(\psi,\theta)$ and $(b,a)$ are respectively measured from the orientation unit vectors for the halo and the disk in the usual way.  

In our analysis, we forced the DM halos into their `edge-on' orientations by setting $\psi = 0$.  In our early simulations in which we allowed $\psi$ to vary, a degeneracy in the parameter space between $Q$, and the inclination of the halo into the sky frustrated our attempts to extract best fit values.  An edge-on orientation was always within the degenerate best-fit region and was always associated with the least-spheroidal (closest to unity) best-fit value of $Q$.  To keep our measurements of Fornax's ellipticity conservative while also facilitating the extraction of best-fit parameters, we restricted our sampled halos to edge-on configurations.  Our measurements of $Q$ should thus be viewed as lower bounds on Fornax's true deviation from sphericity. 




\begin{deluxetable*} { C C C C C C C C  } 
 \tablecaption{Parameters used in the MCMC and bootstrap analyses for the various Fornax distributions with no disk.  \label{tab:WODMCMCChain} }
\tablehead  {\colhead{Analysis Property}  & \colhead{$M (M_{\odot} \times 10^6)$} & \colhead{$r_s$ (\textrm{pc})} & \colhead{$Q^{\dagger}$}                         & \colhead{$\theta^{\dagger}$ (deg)}                 & \colhead{$\psi ^{\dagger}$  }         & \colhead{$\mathbf{h}_c \textrm{ (pc)}$ }  }               
\startdata
 \textrm{Full MCMC Step Size}                       & $10 ^ {1.39}$                                     & $79.1$                               & $0.01^{\dagger}$            & $1.8/\pi^{\dagger}$                               & $\textendash^{\dagger}$           & $(7.91, 7.91, 7.91)$                \\
  \textrm{Full MCMC Range}                         & $[10 ^ {1.39}, 4 \times 10 ^ {3.39}$]                                   & $ [79.1, 7910]$                & $([0.1, 1] \textrm{ or } [1, 10]) ^{\dagger}$           & $[0, \pi ]^{\dagger}$                               & $0^{\ast \dagger}$          & $([-791, 791],\ 0^{\ast},\ [-791, 791])$          \\   
   \textrm{Bootstrap Seed Selection Size}      & $4 \times 10 ^ {2.39}$                    & $395.5$                               & $0.1^{\dagger}$            & $18/\pi^{\dagger}$                             & $\textendash^{\dagger}$           & $(79.1 , 0, \ 79.1)$            
 \enddata
 \tablecomments{We ran a total of 84 chains for each halo type (42 for a prolate distribution and 42 for an oblate distribution), each consisting of 20000 steps.  The $M$, $r_s$, $h_{c,x}$, and $h_{c,z}$ sampling parameters were based on the observed properties of Fornax listed in Table \ref{tab:fornExtras}.  Specifically, we bound $M$ between $0.1$ and $400$ times $M_{\star}$, $r_s$ between $0.1$ and $10$ times $r_{1/2}$, and $h_{c,x}$ and $h_{c,z}$ between $\pm1$ times $r_{1/2}$. }
 \tablenotetext{*}{This parameter was fixed to the shown value}
 \tablenotetext{\dagger}{This parameter was sampled atypically, in the way discussed in Section \ref{sec:MCMC}. }
\end{deluxetable*}

\begin{deluxetable*} { C C C C C C C C  } 
 \tablecaption{Parameters used in the MCMC analyses for the various Fornax distributions when a disk is allowed.  \label{tab:WDMCMCChain}}
\tablehead  {\colhead{Analysis Property}  & \colhead{$\epsilon$}                   & \colhead{$R_d\textrm{ (pc})$}  & \colhead{$\lambda$}       & \colhead{$a \textrm{ (deg)}$ }                   & \colhead{$b \textrm{ (deg)}$  }                  & \colhead{$\mathbf{d}_c \textrm{ (pc)}$ }  }              
\startdata
 \textrm{Full MCMC Step Size}                       & $0.01$                                    &$79.1$                   &    $\textendash^{\dagger}$                         & $\textendash^{\dagger}$    & $\textendash^{\dagger}$   & $(7.91, 7.91, 7.91)$                \\
  \textrm{Full MCMC Range}                         & $[0.0, 1.0]$                        & $[79.1, 7910]$      & $0.1^{\ast}$                    & $178 ^{\ast \dagger}$            & $0^{\ast \dagger}$            & $([-791, 791],\ 0^{\ast},\ [-791, 791])$                    
 \enddata
 \tablecomments{We ran a total of 84 chains for each halo type (42 for a prolate distribution and 42 for an oblate distribution), each consisting of 20000 steps.  The halo parameters $\{M, r_s, h_{x,c}, h_{z,c}\}$ were also sampled, with the parameters shown in Table \ref{tab:WODMCMCChain}.  The $\{Q, \theta\}$ parameters were fixed to their no-disk best fit values.  The ranges and step sizes of $R_d$ and $\mathbf{d}_c$ were respectively set equal to those of $r_s$ and $\mathbf{h}_c$.}
 \tablenotetext{*}{This parameter was fixed to the shown value}
 \tablenotetext{\dagger}{This parameter was sampled atypically, in the way discussed in Section \ref{sec:MCMC}.}
\end{deluxetable*}

We extracted the best fit parameters from the combined chains in various ways, such as by fitting a Gaussian to the projection of the chains onto the axis of a single variable.  We calculated the relative likelihoods of each of the converged to points via the method of Section \ref{sec:compLikelihood}.  We discuss the MCMC results in Sections \ref{sec:fornResWOD} (no disk allowed) and \ref{sec:fornResWD} (a disk allowed).  

\subsubsection{Determining the Distribution of the Measured Ratios} \label{sec:fornBootstrap} 

The MCMC algorithm described in Section \ref{sec:MCMC} allowed us to determine for each DM distribution the parameter values that best match the observed properties of Fornax.  The techniques of Section \ref{sec:compLikelihood} then enabled us to determine which of these best fit models best matches the data by calculating their likelihood ratios.  We determined the significance of the computed likelihood ratios by implementing the bootstrap methodology discussed in Section \ref{sec:measureSig}. 

As outlined by \cite{Efron79}, we simulated random stellar distributions of the dSph under study by randomly selecting with replacement a number of stars equal to the total number of observed stars from the set of observed stars.  We performed a truncated version of the MCMC analysis of Section \ref{sec:MCMC} on this new stellar configuration, shortening the MCMC chains and randomly placing the seeds near the best fit points identified when the large MCMC analysis was run on the true set of stars.  
We chose to shorten the chains and seed them near the previous point of convergence to make the bootstrapping process computationally feasible and with the expectation that the properties of the bootstrapped data should be similar to those of the true data.
By repeating this process many times, we created a distribution of $r_{X,Y}$ values for models $X$ and $Y$ from random manifestations of the same underlying observational surface probability density of Fornax.  It is this distribution of $r_{X,Y}$ that we used to determine if one model is statistically favored.  
 
\section{Results and Discussion} \label{sec:resAndDisc} 

We applied the methodology outlined in Section \ref{sec:theMethod} to Fornax using the analysis steps discussed in Section \ref{sec:FornaxExample}.  Here, we describe and discuss the results of these analyses, first when no disk is considered (Section \ref{sec:fornResWOD}) and then when a disk is considered (Section \ref{sec:fornResWD}). 

\subsection{The Best Fit Spheroidal DM Halo of Fornax} \label{sec:fornResWOD}

The best fit parameters and their uncertainties for prolate and oblate NFW, Acored, and Burkert halos when no disk is considered ($\epsilon = 0$) are listed in Table \ref{tab:fornBestFitWOD}.  The distribution of best fit model ratios for all model pairs are listed in Table \ref{tab:WODLikeRats}. 

\begin{deluxetable*} { C C C C C C C C  } 
 \tablecaption{The best fit parameters for the various proposed DM distributions for Fornax with no disk.  \label{tab:fornBestFitWOD} }
\tablehead  {\colhead{Halo Type}  & \colhead{$M \ (10 ^ 6M_{\odot})$} & \colhead{$r_s$ (\textrm{pc})} & \colhead{$Q$}  & \colhead{$\theta$ (deg) }   & \colhead{$\psi$ (deg)  }   & \colhead{$\mathbf{h}_c \textrm{ (pc, pc, pc)}$ }  & \colhead{$\ln \tilde{\Like}$} }           
 \startdata
 \textrm{NFW; oblate}                    & $9780_{-340}^{+ \ ? \ } $                                               & $7280_{-140}^{+ \ ? \ } $                              & $0.425 \pm 0.032$            & $134.1 \pm 2.3 $                               & $0^{\ast}$           & $(-71 \pm 13 ,\  0^{\ast},\ -143 \pm 12)$                   & 21421.4 \\
  \textrm{NFW; prolate}                 & $9770_{-920}^{+ \ ? \ } $                                                & $ 3920_{-280}^{+ \ ? \ } $                             & $2.17 \pm 0.14$           & $44.2 \pm 2.3 $                               & $0^{\ast}$          & $(-70 \pm 13,\ 0^{\ast},\ -144 \pm 12)$                     & 21421.3 \\        
  \textrm{Acored; oblate}               & $3600 \pm 1300$                                                & $1920 \pm 410$                              & $0.470 \pm 0.029$            & $133.8 \pm 2.1$                                & $0^{\ast}$          & $(-72 \pm 13,\ 0^{\ast},\ -144 \pm 12)$                     & 21430.0 \\  
  \textrm{Acored; prolate}              & $3100 \pm 1400$                                                & $1000 \pm 390 $                                & $2.01 \pm 0.11$          & $43.8 \pm 2.1 $                                & $0^{\ast}$           & $(-73 \pm 12,\ 0^{\ast},\ -144 \pm 11)$                     & 21429.9 \\ 
  \textrm{Burkert; oblate}               &$500 \pm 110$                                                  & $1450 \pm 160 $                              & $0.501 \pm 0.033$           & $133.6 \pm 2.0 $                               & $0^{\ast}$          & $(-74 \pm 13,\ 0^{\ast},\ -145 \pm 12)$                      & 21429.1 \\ 
  \textrm{Burkert; prolate}              &$440 \pm 100 $                                                  & $810 \pm 90 $                                & $1.901 \pm 0.085$           & $43.5 \pm 2.0$                                & $0^{\ast}$          & $(-74 \pm 13,\ 0^{\ast},\ -145 \pm 12 )$                     & 21429.0 \\ 
 \enddata
 \tablecomments{With the exception of $M$ and $r_s$, the measured values and uncertainties were respectively taken to be the mean and width of a Gaussian distribution fitted to the projection of the MCMC chains into the axis of the appropriate parameter.  The uncertainties of $M$ and $r_s$ were determined by fitting a Gaussian in the same way, and their reported values were determined by fitting a curve to the projection of the MCMC chains into the $(M, r_s)$ plane as described in Section \ref{sec:fornMCMCWOD}.  The values of $M$ reported here for the NFW halos are at the edge of the permitted $M$ parameter range.  The true best fit NFW values of $M$ and $r_s$ are almost certainly higher than those reported here, and we do not list their upper uncertainties.   
 }
  \tablenotetext{*}{ This parameter was fixed to the listed value. }
\end{deluxetable*}

The likelihood ratios of Table \ref{tab:WODLikeRats} indicate that Fornax is much better described by a cored halo than by a cusped halo, that Fornax is somewhat better described by an Acored DM halo than by a Burkert DM halo, and that there is no statistically significant preference for either a prolate or an oblate halo. 

Both varieties of cored halo (Burkert and Acored) are strongly preferred over the cusped NFW halo.  We note that, while prolate and oblate Burkert and Acored halos converged within the parameter space of our MCMC, the best fit $M$ values for the prolate and oblate NFW halos lie on the edge of the considered $M$ range.  This suggests that the true best fit NFW halos are likely characterized by larger values of $M$ and $r_s$ than those reported here.  However, the current mass upper limit, listed in Table \ref{tab:WODMCMCChain}, is $400$ times the reported stellar mass of Fornax, listed in Table \ref{tab:fornExtras}.  A larger value of $M$ would push the Fornax mass-to-light ratio above that of the most DM dominated LG dSphs despite the consensus that Fornax is likely one of the least DM dominated such objects.  
We suspect that, in an effort to fit a cusped profile to a seemingly cored DM distribution, our algorithm made the NFW halo as large as possible to minimize the slope of the interior mass density contours.  

\begin{deluxetable*} { C C C  } 
\tablecaption{The bootstrap measurements of the relative likelihood between various DM halos without disks.  \label{tab:WODLikeRats} }
 \tablehead{ \colhead{Ratio type }  & \colhead{Fraction of stellar arrangements } & \colhead{Best fit of bootstrap measured } \\
 \colhead{(always $< 1$ on real data)}  & \colhead{ that return ratio $ > 1$} & \colhead{$\ln$ likelihood ratios}} 
  \startdata 
 \textrm{(NFW-prolate) / (NFW-oblate) }               & $305/1000$ & $-0.07 \pm 0.17$ \\ 
  \textrm{(Acored-prolate) / (Acored-oblate) }             & $474/1000$ & $-0.01 ,\pm 0.21 \\ 
  \textrm{(Burkert-prolate) / (Burkert-oblate) }        & $416/1000$ & $-0.06 \pm 0.27$\\
 \textrm{(NFW-oblate) / (Acored-oblate) }                & $0/1000$ & $-3.78 \pm 0.22$\\ 
 \textrm{ (NFW-prolate) / (Acored-prolate) }              & $0/1000$ & $-3.84 \pm 0.26$\\ 
 \textrm{(NFW-oblate) / (Burkert-oblate) }              & $0/1000$ & $-3.38 \pm 0.28$\\ 
 \textrm{(NFW-prolate) / (Burkert-prolate) }           & $0/1000$ & $-3.40 \pm 0.32$\\ 
 \textrm{ (Burkert-oblate) / (Acored-oblate) }             & $40/1000$ & $-0.40 \pm 0.22$\\ 
 \textrm{(Burkert-prolate) / (Acored-prolate) }          & $51/1000$ & $-0.44 \pm 0.29$ \\ 
 \enddata
  \tablecomments{The distributions of the likelihood ratios are well described by normal distributions, though some are characterized by elongated wings.  Cored halos of either variety are always a better match to the bootstrapped data than any NFW halo.  There is a moderate ($\simeq 2\sigma$) preference for the Acored halo over the Burkert halo.  For all halo varieties, any preference for a prolate or oblate halo can be reversed easily by rerunning the analysis for a different set of the observed stars.}
\end{deluxetable*}

\begin{deluxetable*} { C C C C C C C C } 
\caption{The best fit parameters for the various proposed DM models with a disk.}
\label{tab:fornBestFitWD}
 \tablehead{ \colhead{Halo Type }  & \colhead{$M$} & \colhead{$r_s$} & \colhead{$Q$}  & \colhead{$\theta$}              & \colhead{$\psi$}  & \colhead{$\mathbf{h}_c \textrm{ (pc, pc, pc)}$  }  &  \colhead{$\ln \tilde{\Like}$}\\      
                    \colhead{\textrm{\textemdash}}  & \colhead{$\epsilon$} & \colhead{$R_d \textrm{ (pc)}$} & \colhead{$\lambda$}  & \colhead{$a$}              & \colhead{$b$}  & \colhead{$\mathbf{d}_c \textrm{ (pc, pc, pc)}$}  & \colhead{\textrm{\textemdash}} }                      
\startdata
 \textrm{NFW; oblate}                      & $9700 \pm 1600$ & $7100 \pm 1600$                    & $0.427^{\ast}$                & $134.1^{\ast}$                          & $0^{\ast}$                              & $(-77 \pm 28 , 0^{\ast},-138 \pm 28  )$  & 21421.8\\
  \textrm{\textemdash}                      & $0.030_{-0.019}^{+0.203}$ & $6000 \pm 1300$                 & $0.1^{\ast}$                & $178 ^ {\ast}  $                          & $0^{\ast }$                              & $(-6.0 \pm 670 , 0^{\ast},-370 \pm 420  )$ & \textrm{\textemdash}  \\ 
 \textrm{NFW; prolate}                     & $9700 \pm 1600$ & $3900 \pm 1300$                    & $2.16^{\ast}$                & $44.2^{\ast }$                          & $0^{\ast}$                              & $(-76 \pm 29 , 0^{\ast},-137 \pm 26  )$  & 21421.6\\
  \textrm{\textemdash}                     & $0.039_{-0.024}^{+0.143}$    & $6000 \pm 1300 $               & $0.1^{\ast}$                & $178 ^ {\ast} $                           & $0^{\ast}$                             & $(-80 \pm 250 , 0^{\ast},-350 \pm 300 )$ & \textrm{\textemdash} \\
 \textrm{Acored; oblate}                     & $3600 \pm 1000$ & $1900 \pm 320$                    & $0.466^{\ast }$                & $133.7^{\ast}$                          & $0^{\ast}$                           & $(-73 \pm 17 , 0^{\ast},-144 \pm 16  )$  & 21430.0\\  
 \textrm{\textemdash}                     & $0_{-0}^{+0.14}$                            & $6800 \pm 1100$ & $0.1^{\ast}$                & $178 ^ {\ast}$                           & $0^{\ast}$                             & $(-120 \pm 70, 0^{\ast}, 220 \pm 410)$ & \textrm{\textemdash}\\
 \textrm{Acored; prolate}                   & $2870 \pm 880 $ & $1000 \pm 170$                  & $2.04^{\ast r}$                            & $43.8^{\ast}$ & $0^{\ast}$                                              & $(-74 \pm 17 , 0^{\ast},-146 \pm 16  )$ & 21424.9 \\ 
  \textrm{\textemdash}                  & $0_{-0}^{+0.14}$                            & $1290 \pm 260$ & $0.1^{\ast}$                & $178 ^ {\ast } $                           & $0^{\ast}$                              & $(-86 \pm 24, 0^{\ast}, -150 \pm 14  )$ & \textrm{\textemdash} \\ 
  \textrm{Burkert; oblate}                 & $390 \pm 770$ & $1290 \pm 260$                 & $0.466^{\ast}$                             & $133.7^{\ast}$ & $0^{\ast}$                                & $(-73 \pm 16, 0^{\ast},-147\pm 16  )$ & 21428.6 \\ 
    \textrm{\textemdash}                 & $0.0041_{-0.0013}^{+0.0028}$                             & $8100 \pm 2200 $ & $0.1^{\ast}$                & $178 ^ {\ast}$                          & $0^{\ast }$                              & $(-180 \pm 280 , 0^{\ast}, \textrm{\textemdash}^ {\ddagger}   )$ & \textrm{\textemdash}\\
 \textrm{Burkert; prolate}                & $520 \pm 370$ & $840 \pm 130$                 & $2.02^{\ast }$                             & $43.9^{\ast }$ & $0^{\ast }$                                & $(-72 \pm 18 , 0^{\ast},-145 \pm 15  )$  & 21428.4  \\ 
  \textrm{\textemdash}               & $0.0040_{-0.0013}^{+0.0036}$            & $8300 \pm 1700$ & $0.1^{\ast}$  & $178 ^ {\ast} $ & $0^{\ast}$  & $(-170 \pm 23 , 0^{\ast},-140 \pm 78  )$ & \textrm{\textemdash}\\ 
 \enddata
 \tablecomments{Some of the halo parameter values were fixed to the best fit values in Table \ref{tab:fornBestFitWOD}.  The values and uncertainties were respectively equal to the mean and width of fitted Gaussian distributions, though the MCMC results were often poorly described by a Gaussian fit.  As with the measurement of the halo parameters for the configurations without a disk, the values of $M$ and of $r_s$ were determined by fitting a curve to the projection of the MCMC chains onto the $(M, r_s)$ plane, and their uncertainties were determined by fitting separate Gaussian distributions to the projection of the MCMC chains onto the $M$ and $r_s$ axes.  We fit $\epsilon$ with a beta distribution \citep{James08}.  We report the mode of the fitted beta distribution as the best fit value of $\epsilon$, and the upper and lower bounds as the left and right distances from the mode that encompass the same portions of the probability distribution as the portion of a normal distribution enclosed by the mean and one standard deviation from the mean.}
\tablenotetext{*} {This parameter was fixed to the listed value. }
   \tablenotetext{\ddagger}{The parameter failed to clearly converge and our Gaussian fitting method failed.}
\end{deluxetable*} 

We further find that the Fornax DM halo is characterized by several well-defined morphological parameters for all considered DM halos.  We consistently measured $\theta$, equivalent to the position angle of Fornax's symmetry axis measured north to east when the halos are edge on, to be about $134^{\circ}$ for oblate halos and $44^{\circ}$ for prolate halos, with little inter-halo variation.  The prolate value is close to the roughly $40^{\circ}$ to $45^{\circ}$ position angles of Fornax reported by \cite{Walker11, McConnachie12, delPino15, Munoz18}.  

We measured the center of Fornax's DM halo, $\mathbf{h_c}$, to be about $(-70 \textrm{ pc}, 0 \textrm{ pc}, -140 \textrm{ pc})$ for all halo types, where recall the separation along the axis of the viewer, $ h_{c,y}$, is not varied.  In sky coordinates, this value of $\mathbf{h_c}$ places Fornax's DM center of mass at $(\textrm{R. A. (J2000), Decl. (J2000)}) = ((02 \textrm{h}: 39\textrm{m}, : 51\textrm{s}),(-34\textrm{d}: 30\textrm{m}, : 16\textrm{s}))$, within the $1\sigma$ limits of the photometric center of Fornax reported by \cite{Walker11, McConnachie12, delPino15, Munoz18}.  

We measured Fornax's semi-minor to semi-major axis ratio to be about $0.45$, with the NFW DM halo having a slightly smaller value.  This is significantly more spheroidal than the measurement of Fornax's photometric axis ratio of about $0.7$ reported by \cite{McConnachie12, delPino15, Munoz18}.  This is to be expected.  A spheroidal distribution of matter sources a more spherical gravitational potential.   

Our measured ellipticities for the cusped and cored DM halos are, respectively, consistent with and slightly less extreme than the corresponding ellipticities measured by \cite{Hayashi12}.  However, unlike \cite{Hayashi12}, we find that both prolate and oblate halos of comparable ellipticities are similarly consistent with the data.  

These measured ellipticities are at the extreme end of the DM halo ellipticities predicted by the $\Lambda$CDM with baryon simulations of the FIRE project (\cite{Fitts17} and references therein).  As shown in Figure 4a of \cite{Xu19}, the FIRE simulations predict that less than one in five LG dSphs would have DM axis ratios as large or larger than the DM ellipticity of Fornax we measure.  Further, we assumed that Fornax is viewed edge on.  Therefore, our measurement of Fornax's axis ratio is conservative.  If Fornax is not edge on with respect to an Earth-based observer, it is even more spheroidal.

Summarizing these results, we determined that, when the velocity dispersion is assumed isotropic and each population is characterized by a single bulk rotation, the Fornax LG dSph is much better characterized by a cored spheroidal DM halo than by a cusped spheroidal DM halo.  This spheroidal halo could be either prolate or oblate, and is characterized by an axis ratio at the extreme end of what is predicted by CDM simulations with baryonic feedback.  We discuss the significance of these results in Section \ref{sec:conclusion}. 

In the following Sections, we describe how we arrived at the results of Tables \ref{tab:fornBestFitWOD} and \ref{tab:WODLikeRats} using the methodology and data of Sections \ref{sec:theMethod} and \ref{sec:FornaxExample}. 

\subsubsection{Determining the Best Fit Parameter Values for Halos Without a Disk} \label{sec:fornMCMCWOD}
\begin{figure*} 
    \centering
        \includegraphics[width=1.0\textwidth]{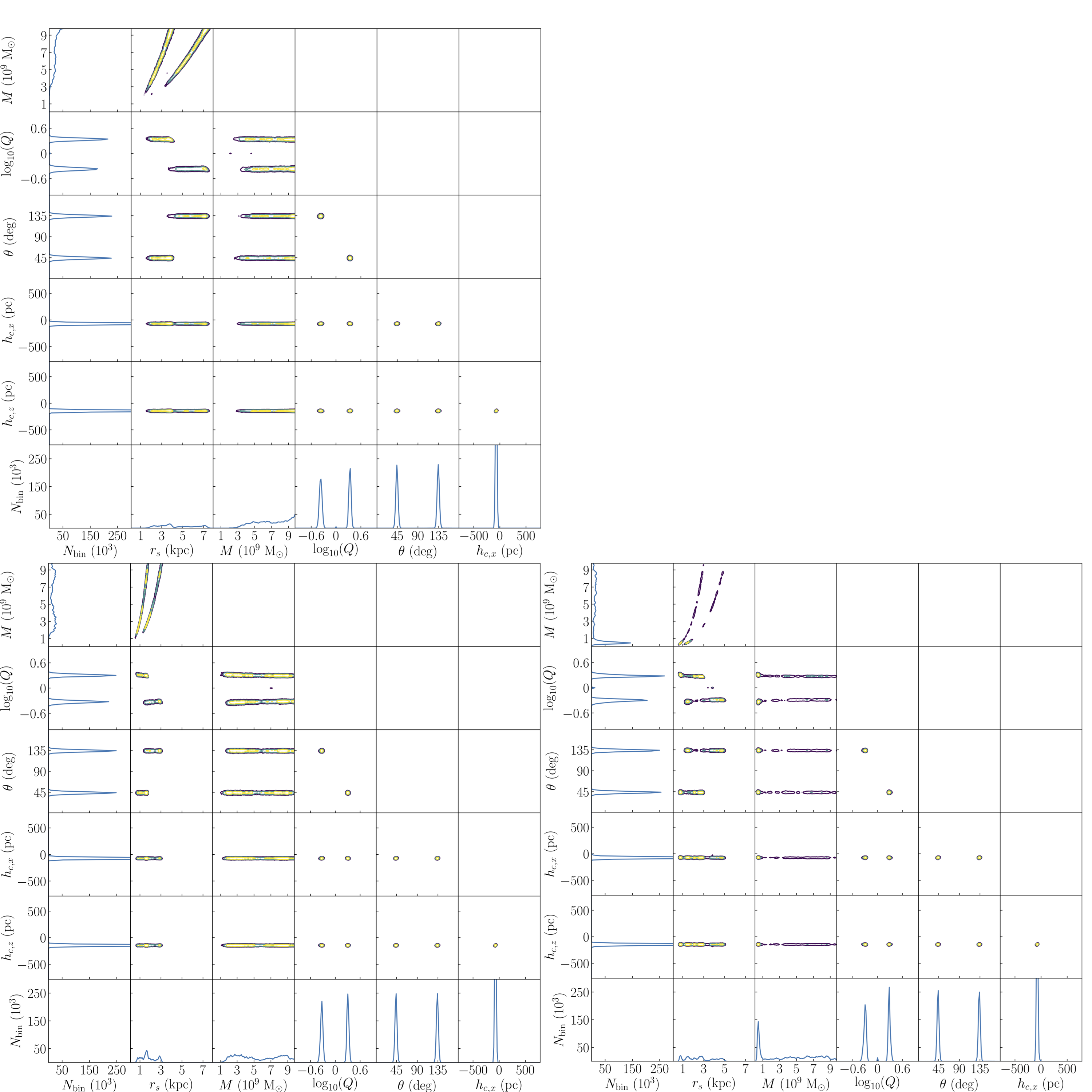}
            \caption{The MCMC results for the NFW (top), Acored (bottom left), and Burkert (bottom right) halos without a disk.  From these MCMC results, we measured the best-fit halo parameters using the methods of Section \ref{sec:fornMCMCWOD}, and report them in Table \ref{tab:fornBestFitWOD}.  There are two clear regions of convergence, reflecting the physical bifurcation of the halo parameter space into prolate ($Q>1$) and oblate ($Q < 1$) sections.  For the NFW halo, the chains converged to the edge of the allowed parameter space along the $M$-axis, indicating that the NFW DM halo that best matches the data is more massive than the MCMC upper mass limit listed Table \ref{tab:WODMCMCChain}.  In an attempt to match a cusped DM halo to a stellar distribution that is more naturally described by a cored halo, our algorithm enlarged the mass and size of the NFW halo to reduce the slope of its interior mass contours.  The MCMC chains converged to halo parameters that are well within the considered parameter spaces for the Acored and Burkert models.} \label{fig:MCMCNWOD}
\end{figure*}

The specific parameters used in the final MCMC chains without a disk are listed in Table \ref{tab:WODMCMCChain} and their outputs for the NFW, Burkert, and Acored halos are shown in Figure \ref{fig:MCMCNWOD}.  We show both the prolate and oblate chains in the same set of plots. 

The $Q$, $\theta$, $h_{c,x}$, and $h_{c,z}$ variables all converged to two well localized points in each parameter space: one for prolate and on for oblate DM halos.  Because both regions are of physical significance, we forced half of our chains to sample only the prolate space and half to sample only the oblate space.  

The best fit values of $Q$, $\theta$, $h_{c,x}$, and $h_{c,z}$ were determined by fitting a Gaussian curve of variable height, width, and central value to the projection of the MCMC results onto the axis of each of these parameters.  The projections of the combined prolate and oblate chains are shown in the bottom most windows and left most windows of each of the subfigures in Figure \ref{fig:MCMCNWOD}.  The central values and widths of these Gaussian fits were taken respectively as the best fit values and the uncertainty in the best-fit values for each of the halo models, given the stellar data of Fornax.  These values are listed in Table \ref{tab:fornBestFitWOD}. 

The $M$ and $r_s$ parameters (total halo mass and scale radius) did not clearly converge to localized points in the parameter space.  Rather, each halo type has two convergence regions, one for prolate and one for oblate halos, through the $M$ and $r_s$ parameter spaces.  These extensive regions are portions of the parameter space in which the correspondence between the halo configurations and the observed stars varies relatively slowly.  To determine the best fit values $M$ and $r_s$, we fixed $Q$, $\theta$, $h_{c,x}$, and $h_{c,z}$  to their measured best fit values and explicitly computed the likelihoods for $(M,r_s)$ pairs along the convergence curves.  We determined the best fit values of $M$ and $r_s$ by determining the point along the $(M,r_s)$ convergence curve that maximizes the correspondence between the halo and the observed stars.  We estimated the uncertainties of these best fit values by fitting Gaussian distributions to the one dimensional projections of the MCMC chains into the axes of each parameter and taking the widths of the fitted distributions.  These values are listed in Table \ref{tab:fornBestFitWOD}. 


The best fit values acquired from this MCMC analysis are the parameters with which a particular halo model best matches the Fornax data.
We also used these parameters as initial guesses with which we seed the bootstrap analysis.

\subsubsection{Determining the Significance of the Relative Likelihoods of Each Halo} \label{sec:fornLikelihoodWOD}
We ran the bootstrap method of Section \ref{sec:fornBootstrap} $1000$ times.  The centers of the seed regions are the best fit values listed in Table \ref{tab:fornBestFitWOD} and the parameter box sizes in which the seeds were randomly placed are listed in Table \ref{tab:WODMCMCChain}.  The best fit parameters for each bootstrap were measured by an automated version the same methods outlined in this Section, starting with running individual MCMC chains and ending with measuring the likelihood along the $(M, r_s)$ convergence regions.  

The relative likelihoods between each of the considered DM halo models measured from the best fit parameters on the Fornax data are listed in Table \ref{tab:WODLikeRats}.  We measured and show in Table \ref{tab:WODLikeRats} the fraction of bootstrap resamplings that flip which of model $X$ or $Y$ best matches the data, along with the approximate value and uncertainty for $r_{X,Y}$ derived from the bootstrap results.  If model $X$ matched the data better than model $Y$ for $10\%$ of resamplings, then there is a roughly $10\%$ chance that the seeming preference in the data for model $Y$ is due only to the particular selection of observable stars in Fornax.  

\subsection{Results of the MCMC on Fornax with a Possible Disk} \label{sec:fornResWD} 

We report the best-fit results when a nonzero disk is allowed in Table \ref{tab:fornBestFitWD}.

Only for the NFW halo did some of the MCMC chains converge to a nonzero value of $\epsilon$ larger than the MCMC step size of one percent.  This preference for the addition of a small disk in the interior of the NFW halo is simply further evidence of the poor correspondence between the cusped NFW halo and the data.
Both the nature of the disk preferred by the NFW MCMC chains and the relative likelihoods of the configurations without a disk and with a disk (Tables \ref{tab:fornBestFitWOD} and \ref{tab:fornBestFitWD}, respectively) support this contention.  The MCMC chains with an NFW halo and a disk converged to a large value of $R_d$, spreading the interior contours of the predicted stellar density profiles.  Thus, the MCMC used the extra degrees of freedom furnished by the disk parameters to make the NFW profiles less cusped.  For an NFW halo, the best fit configuration with a disk matches the Fornax data marginally better than the best fit NFW configuration without a disk.  However, all considered versions of the cored halos match the Fornax data better the cusped NFW halo with or without as disk. 

Because the addition of a disk does not meaningfully improve the correspondence between the cored DM halos and the data, and because the cusped NFW halo with a disk does not match the data as well as any considered cored DM halo, we conclude that there is no evidence of a disk with total mass larger than one percent of Fornax's total mass.  We did not run a bootstrap analysis on these DM models. 

In the following Sections, we summarize the simplifying restrictions that we made and the fitting techniques that we used to extract the best fit parameters listed in Table \ref{tab:fornBestFitWD}. 

\subsubsection{Simplifying the Parameter Space} 
\begin{figure*} 
    \centering
        \includegraphics[width=0.8\textwidth]{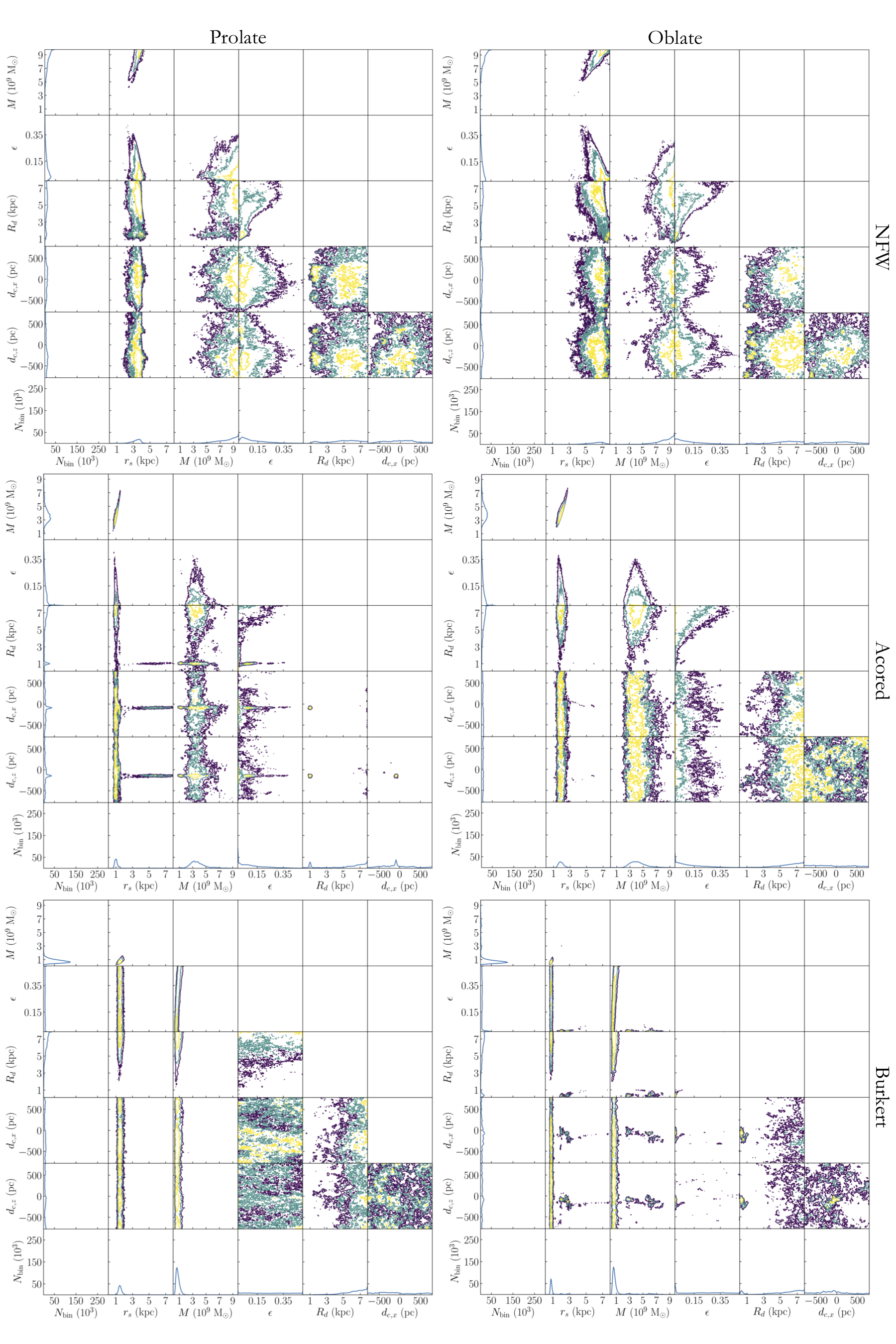}
            \caption{The MCMC results for prolate (left column) and oblate (right column) NFW (top row), Acored (middle row), and Burkert (bottom row) DM halos when the possibility of a disk is allowed ($\epsilon \geq 0$). 
  As in the case without a disk (Figure \ref{fig:MCMCNWOD}), the NFW MCMC chains converged to the edge of the allowed parameter space along the $M$-axis, indicating that the best fit NFW halo has a larger mass than the MCMC's upper mass limit.  The nonzero convergence value of $\epsilon$ and the large convergence value of $R_d$ in the top row suggest that placing a small portion of the total galaxy mass in an extensive disk improves the correspondence between the NFW model and Fornax's RG stars.  The disk spreads the interior contours of the projected stellar number density, flattening the slope of the mass density of the cusped NFW halo.  For the cored halos (middle and bottom rows), the converged to value of $\epsilon$ is functionally 0, indicating that their correspondence to the data is not improved with the addition of a disk.}  \label{fig:MCMCNOWD}
\end{figure*} 

When all parameters in the complete parameter set, $\{M, r_s, Q,\theta, \psi, h_{c,x}, h_{c,z}\, \epsilon, R_d, \lambda, a, b, h_{d,x}, h_{d,z} \}$ were allowed to vary, the results were too complex to easily identify convergence peaks or convergence regions.  We simplified the examined parameter space by fixing a subset of the parameters.  Specifically, we forced both the disk and halo to their `edge-on' orientations ($\psi = 0$ and $b=0$).  Under the assumption that the inclusion of a disk would not impact the apparent global properties of the DM halo, we also set $Q$ and $\theta$ to one of the six identified best fit values listed in Table \ref{tab:fornBestFitWOD}.  We found that the results were largely insensitive to the flatness parameter of the disk, $\lambda$, and we set $\lambda = 0.1$.  
The flattened structure of young main sequence stars identified by \cite{delPino15} provides a natural and meaningful prior value for the orientation of a possible disk, and we fixed $a = 178 ^ {\circ} $ to match the reported orientation of this flattened structure.  

Once the parameters were fixed, we ran the MCMC analysis on the remaining free parameters, $\{M, r_s, h_{c, x}, $ $h_{c,z}, \epsilon, R_d, h_{d, x}, h_{d,z} \}$, for both the prolate and oblate configurations of the three halo types.  We show the results of these chains in Figure \ref{fig:MCMCNOWD}.

\subsubsection{Extracting the Best Fit Parameters} 

We used the methods described in Section \ref{sec:fornResWD}, including the method of simultaneously fitting $M$ and $r_s$, to measure the convergence values of the varied parameters.  However, because the disk is assumed to be much less massive than the halo ($\epsilon \simeq 0$), the other halo parameters, $\{R_d, h_{d, x}, h_{d,z} \}$, often had very little impact on the model fit and were sometimes poorly constrained.  

The $\epsilon$ posterior distributions are asymmetric, and are poorly described by a normal distribution.  The beta distribution, described in \cite{James08}, is a bounded and asymmetric probability distribution that is useful for modeling the posterior distributions of parameters sampled in bounded spaces.  We used a beta distribution to model the $\epsilon$ posterior distributions.  We report the mode of the beta distribution as the best fit $\epsilon$ value, and the left and right $\epsilon$ ranges that contain the equivalent of $\pm 1 \sigma$ of probability in a standard normal distribution as the lower and upper $\epsilon$ bounds.  When $\epsilon$ is measured to be functionally 0, the lower bound is set to $0$.  

We report the convergence results, as they were best measured, in Table \ref{tab:fornBestFitWD}. 

\section{Discussion and Conclusion} \label{sec:conclusion}
 We have developed a technique for probing the substructure of dark matter (DM) halos of Local Group dwarf spheroidal galaxies (LG dSphs) by checking the correspondence between stellar positions and the probability distribution predicted from the Jeans equations given an assumed DM profile and an observationally motivated kinematical model.  Of particular note, our method tests the consistency of an assumed DM distribution with astronomical observations with no reference to an assumed stellar profile.  This technique allowed us to constrain unusual DM structure predicted by non-standard DM models.  

Applying this technique to the red giant (RG) stars in Fornax, we measured the best fit parameters for several spheroidal DM halo models.  Those parameters are listed in Table \ref{tab:fornBestFitWOD}.

Our results suggest that the spheroidal DM halo of Fornax is cored and has a semi-minor to semi-major axis ratio of at least 0.5.  

Specifically, we found that the best-fit cored and cusped DM halo models are significantly spheroidal, with the $2\sigma$ limits on the spheroidal axes ratio somewhere between $0.37$ and $0.58$ for every considered model.  This result suggests that Fornax is more spheroidal than at least eighty percent of dSph halos predicted from the FIRE simulations, according to the the analysis of \cite{Xu19}, even with the conservative assumption that Fornax is viewed `edge-on'.

Comparing distinct best fit halo models, we found that either of two types of cored halos are a much better match to the Fornax data than a cusped NFW halo.  We also found that, for a given halo model, prolate and oblate DM halos describe the data equally well.  

Fornax is one of the most massive LG dSphs in terms of both total stellar and dynamical mass, and is thus one of the LG dSphs most likely to lose an initial CDM cusp to baryonic feedback effects.  Although the general predictions of \cite{Penarrubia12, GarrisonKimmel13} argue that it is of a mass that would make the erasure of a DM cusp by baryonic feedback challenging, the removal of a DM cusp in a Fornax like LG dSph is precisely what the simulations of \cite{Amorisco14, Onorbe15} predict.  The existence of a cored DM halo in Fornax would provide insight into the accuracy of these predictions. 

We also looked for evidence of a disk-like structure within the larger DM halo.  The best fit parameters of this search are listed in Table  \ref{tab:fornBestFitWD}.  

We found that the inclusion of a disk marginally improved the correspondence between the predicted profile of an NFW halo and the observed data by flattening the total interior mass distribution.  The insertion of a disk did nothing to improve the correspondence between the profiles predicted from cored halos and the observed stars.  Disk-free cored halos remained much better better matches to the data than a cusped halo even with the insertion of a disk.  
The lack of evidence of a disk with mass larger than one percent of the total halo mass of Fornax constrains both the possible size of the flattened stellar structure in Fornax and the existence of a dark disk within Fornax's larger DM halo.  

The possibility that one or more of the assumptions and simplifications that we made in the course of our calculation are affecting our measurements does exist, and our results should be understood in that context.  With that caveat noted, our conclusions that Fornax's DM halo is likely both cored and characterized by an axis ratio of at least 0.5 are in mild tension with current predictions of the canonical $\Lambda$CDM cosmology.  Perhaps Fornax is a slight outlier amongst LG dSphs, and future analyses will determine that the population as a whole conforms with the predictions of the canonical cosmology.  Or perhaps Fornax shares its morphological properties with the broader population of the LG dSphs.  In that case, the population of LG dSphs would differ in a systematic way from the present predictions of the $\Lambda$CDM hypothesis, even when baryonic effects are included.  This question can only be answered by the acquisition and analysis of additional LG dSph data. 

Owing to the concerted effort of observational astronomers, including  \cite{Kleyna02, Tolstoy04, Coleman05, Munoz05, Koch06, Battaglia06, Koch07, Martin07, Simon07, Koch08, Battaglia08, Walker09, Walker09_data, Kirby10, Ural10, Battaglia11, Majewski13, Hendricks14, Walker15, Caldwell17, Kacharov17, Gonzalez18}, and to the massive set of astrometric and kinematical data furnished by the GAIA Space Telescope \citep{Gaia16}, the set of dSphs with kinematical and astrometric data is large and growing.  The Large Synoptic Survey Telescope will identify new spheroidal and ultrafaint Local Group dwarf galaxies \citep{Bechtol19, Ando19}, and continuously improving spectroscopic methods with large telescopes will enhance our understanding of the distinct stellar populations within LG dSphs \citep{Ji19}.  A robust study of the properties of these DM dominated objects will offer unparalleled insight into the small scale structure of DM subhalos.  We hope that our technique of using stellar structure to probe DM structure can supplement conventional modeling methods to help provide new insights into the nature of dSph halos, and thus into the nature of the enigmatic substance that is dark matter.

\section{Acknowledgements}
We thank Mathew Walker for his guidance in developing the technique discussed in Section \ref{sec:compLikelihood} and for several helpful and interesting discussions.  We thank Andreas Burkert for his helpful critiques.  We thank Doug Finkbeiner for his advice on refining some of the sampling techniques in the MCMC algorithm discussed for particular parameters.  We thank Linda Xu and Jakub Schultz for their help in deriving the gravitational potentials of spheroidal DM halos discussed in Appendices  \ref{app:haloPot} and \ref{app:diskPot}.  


\bibliography{LisaProjectBibliography}

\appendix

\section{Stellar Densities in an Axisymmetric Gravitational Potential} \label{app:axisymPot}

In this Appendix, we derive an analytic equation for the 3d stellar density, $\nu$, for an axisymmetric system with an isotropic velocity dispersion and fixed rotational velocity in the region of interest.  We thus set out to solve the Jeans equations for a spheroidal symmetric galaxy (Equations 4-29a-4-29c of \cite{BandT}): 
\begin{equation} \label{eq:jEqDerR1}
\frac{\partial \nu \overline{v}_R}{\partial t} + \frac{\partial \nu \overline{v_R^2}}{\partial R} + \frac{\partial \nu \overline{v_R v_z}}{\partial z} + \nu(\frac{\overline{v^2_R} - \overline{v^2_z}}{R} + \frac{\partial \Phi}{\partial R}) = 0 \ ,
\end{equation} 
\begin{equation}
\frac{\partial \nu \overline{v}_z}{\partial t} + \frac{\partial \nu \overline{v_R v_z}}{\partial R} + \frac{\partial \nu \overline{v_\phi v_z}}{\partial z} + \frac{2\nu}{R}\overline{v_\phi v_r}  = 0 \ ,
\end{equation} 
\begin{equation}\label{eq:jEqDerz1}
\frac{\partial \nu \overline{v}_z}{\partial t} + \frac{\partial \nu \overline{v_R v_z}}{\partial R} + \frac{\partial \nu \overline{v_z^2}}{\partial z} + \frac{\nu\overline{v_R v_z}}{R} + \nu  \frac{\partial{\Phi}}{{\partial z}}  = 0 \ ,
\end{equation} 
where we have used the standard cylindrical coordinates.  

In order to simplify these equations to manageable forms, we must make some assumptions about the kinematical information of the dSph of interest.  First, we assume that the distribution has reached equilibrium and thus that all time derivatives are 0.  Further, we assume that the velocity dispersion is isotropic:
\begin{equation} 
\sigma_{ij}^2 = \sigma ^2 \delta_{ij} 
\ ,
\end{equation} 
 and that velocity streaming can occur only in the $\hat{\phi}$ direction:
\begin{equation}
  \overline{v}_i \overline{v}_j = 
  \begin{cases} 
      \overline{v}_\phi ^2 & i = j = \phi  \\
      0 & \textrm{else} 
   \end{cases} \ .  
\end{equation} 
{These assumptions simplify} many of the kinematical terms:
\beq
   \overline{v_\phi^2} &= \sigma_{\phi \phi} ^2 + \overline{v}_\phi ^ 2 = \sigma ^2 + \overline{v}_\phi ^ 2 \\ \ ,
   \overline{v_z^2} &= \sigma_{zz}^2 + \overline{v}_z ^2 = \sigma_{zz}^2 + 0 =\sigma ^2 \\ \ ,
 \overline{v_R^2} &= \sigma_{RR}^2 + \overline{v}_R ^2 = \sigma_{RR}^2 + 0 =\sigma ^2   \ , 
\eeq
 thus reducing Equations \ref{eq:jEqDerR1} and \ref{eq:jEqDerz1} to 
\begin{equation} \label{eq:jEqDerR2}
\frac{\partial \nu(R,z) \sigma^2(R,z)}{\partial R} - \nu(R,z) (\frac{\overline{v}_\phi^2(R,z)}{R} - \frac{\partial \Phi(R,z)}{\partial R} )= 0 \ ,
\end{equation}
\begin{equation} \label{eq:jEqDerz2}
\frac{\partial \nu(R,z) \sigma^2(R,z)}{\partial z} + \nu(R,z) \frac{\partial \Phi(R,z)}{\partial z} = 0 \ .
\end{equation}

We must now propose analytic expressions for both $\sigma^2(R,z)$ and $\overline{v}_\phi(R,z)$.  For simplicity, we assume that $\sigma^2(R,z)$ is effectively constant over the region of interest.  However, we again emphasize that our method can handle more complex forms of $\sigma^2(R,z)$.  Such changes would just alter the final number of kinematical parameters, and the relation between $\nu$, $\Phi$, and the kinematical parameters in Equation \ref{eq:fullPotFunct}.  We allow each population to exhibit a single, overall rotation term by setting $\overline{v}_\phi(R,z)$ = $\omega R$, where $te\omega$ is the fixed angular velocity of the dSph population.  

For the specific forms of these kinematical terms, Equations \ref{eq:jEqDerR2} and \ref{eq:jEqDerz2} reduce to the solvable forms:
\beq \label{eq:jEqDerR3}
& \sigma ^ 2 \frac{\partial \nu(R,z) }{\partial R} - \nu(R,z) (\omega^2 R - \frac{\partial \Phi(R,z)}{\partial R} )= 0 
& \Rightarrow \frac{\partial \ln{\nu(R,z)} }{\partial R} - \frac{1}{\sigma^2} (\omega^2 R - \frac{\partial \Phi(R,z)}{\partial R} )= 0 \ ,
\eeq
\beq \label{eq:jEqDerz3}
& \sigma ^ 2 \frac{\partial \nu(R,z)}{\partial z} + \nu(R,z) \frac{\partial \Phi(R,z)}{\partial z} = 0 
&\Rightarrow \ln{\nu(R,z)} = F(R) - \frac{1}{\sigma^2} \Phi(R,z) \ ,
\eeq
for some arbitrary function $F(R)$.  

Combining Equations \ref{eq:jEqDerR3} and \ref{eq:jEqDerz3}, we find that
\beq
&\frac{\partial (F(R) - \frac{1}{\sigma^2} \Phi(R,z) ) }{\partial R} - \frac{1}{\sigma^2} (\omega^2 R - \frac{\partial \Phi(R,z)}{\partial R} )= 0 
&\Rightarrow \frac{d F(R)}{d R} - \frac{\omega^2 R}{\sigma^2 } = 0 
& \Rightarrow F(R) = \frac{\omega^2 R^2}{2 \sigma ^2} + D \ ,
\eeq
for some constants $R_0$ and $D$.  Thus, 
\beq     \label{eq:JEqSol}
\nu(R,z) &= C \exp{( \frac{\omega^2 R^2}{2 \sigma ^2} - \frac{\Phi(R,z)}{\sigma^2} )} \ ,
\eeq
where $C$ is a normalization constant with units of (distance)$^{-3} $.  

Equation \ref{eq:JEqSol} gives us an analytic relation between the three-dimensional spatial probability distribution of a stellar population, $\nu$, the two characteristic kinematical parameters of the population, $\sigma^2$ and $\omega$, and the ambient gravitational potential of the dSph, $\Phi$.  For our example, we measure $\sigma^2$ and $\omega$ from observational data (see Section \ref{sec:data}) and derive $\Phi$ from an assumed DM distribution (see Section \ref{sec:DDDM}).  

\section{Stellar Densities in a Spherically Symmetric Gravitational Potential} \label{app:sphericalPot}

In this Appendix, we derive a differential equation for the 3d stellar density, $\nu$, for a spherical system with an anisotropic velocity dispersion and no rotation.  We thus set out to solve the Jeans equations for a spherically symmetric galaxy (Equations 4-30 of \cite{BandT}): 
\begin{equation} \label{eq:jEqDerSph1}
 \frac{d\nu \overline{v_r^2}}{dr} + \frac{\nu}{r} \Big [ 2 \overline{v_r^2} - (\overline{v_\theta^2} + \overline{v_\phi^2}) \Big ] + \nu \frac{d \Phi}{d r} = 0 \ ,
\end{equation} 
where we have used standard spherical coordinates.  

If the system is spherically symmetric {in kinematics and morphology}, than:
\beq
\sigma_R ^2 & =  \overline{v_R^2}  \\
\sigma_\phi ^2 &= \overline{v_\phi^2} = \sigma_\theta ^2 = \overline{v_\theta^2} \ ,
\eeq
and the kinematics are fully understood once the $r$ dependence of $\sigma_r ^2$ and $\sigma_\phi ^2$ are specified.  We use the anisotropy of the system's velocity ellipsoidal as a function of $r$, 
\beq
\beta(r) = 1 - \frac{\sigma_\theta^2 }{\sigma_r^2}    \ ,
\eeq
to rewrite Equation \ref{eq:jEqDerSph1} in terms of the three to-be-specified functions, $\Phi(r)$, $\sigma^2_r(r)$, and $\beta(r)$: 
\begin{equation} \label{eq:JEqDerSph2} 
  \sigma_r^2(r) \frac{d\nu(r)}{dr} + \frac{d \sigma_r^2(r)}{dr} \nu(r)+ \frac{2 \nu(r) \sigma_r^2(r) \beta(r)}{r} + \nu(r) \frac{d \Phi(r)}{d r} = 0 \ .
\end{equation} 
We now isolate $\nu(r)$ and integrate: 
\beq \label{eq:JEqDerSph3} 
  \int _{r_0} ^ {r} dr' \frac{1}{\nu(r)} \frac{d\nu(r)}{dr} & =  - \int _{r_0} ^ {r} dr'  \Big (\frac{1}{\sigma_r^2} \frac{d \sigma_r^2(r)}{dr} + \frac{2 \beta(r)}{r}  + \frac{1}{\sigma_r^2(r)} \frac{d \Phi(r)}{d r}  \Big ) \\ 
\Rightarrow  \nu(r) & = \nu(r_0) \frac{\sigma_r^2(r_0)}{\sigma_r^2(r)} \exp{\Big (- \int _{r_0} ^ {r} dr' \Big ( \frac{2 \beta(r)}{r}  + \frac{1}{\sigma_r^2(r)} \frac{d \Phi(r)}{d r}  \Big ) \Big)} \ . 
\eeq  
Once the precise forms of the three relevant functions are specified, the integrals in Equation \ref{eq:JEqDerSph3} can be computed and an analytic relation between $\nu$ and the gravitational potential and kinematics of the system thus specified.  

\section{Computing $\Phi_{h}$} \label{app:haloPot} 

We here derive an expression for the potential of a halo with a mass distribution, $\rho_{h}$, that has some specified interior mass density, $\rho_{\textrm{in}} $, up to some spheroidal cutoff radius, $r_c = c r_s$, at which $\rho_h$ discontinuously falls to $0$: 
\begin{equation} \label{eq:genDMDens} 
    \rho_{h}(\mu) = 
    \begin{cases}
        \rho_{\textrm{in}} (\mu) & \mu \leq c \\
        0 & \mu > c \ ,
    \end{cases}
\end{equation}
where we define $m \equiv \sqrt{R^2 + (z/Q) ^2}$ for ellipticity $Q$ and $\mu \equiv m/r_s$.  

We now solve for $\Phi_{h}$ from $\rho_{h}$.  According to Equation (2-99) of B\&T, if a mass density, $\rho$, is characterized by triaxial, equidensity surfaces with spheroidal radii $m^2 \equiv a_1 ^ 2 \sum_{i = 1} ^ 3 x_i ^ 2/a_i^2$, the the gravitational potential, $\Phi$, of $\rho$ is given by 
\begin{equation} \label{eq:BandTElPhi}
    \Phi(\mathbf{x}) = - \pi G (\frac{a_2 a_3}{a_1}) \int_0 ^ {\infty} d \tau \frac{\psi(\infty) - \psi(\tilde{m}(\mathbf{x}, \tau))}{\sqrt{(\tau + a_1^2)(\tau + a_2^2)(\tau + a_3^2)}} \ , 
\end{equation}
where $a_1$, $a_2$, and $a_3$ are constant shape parameters, $\psi(\tilde{m}) \equiv \int_0 ^ {\tilde{m}^2} d (m'^2) \rho(m'^2) = 2 \int_0 ^ {\tilde{m}} d m' m' \rho(m')$, and $\tilde{m} (\mathbf{x},\tau)^2 = a_1 ^ 2 \sum_{i = 1} ^ 3 (x_i ^ 2)/(a_i^2 + \tau)$. 

In our case, $a_1 = a_2 = 1$, $a_3 = Q$ and $\rho(m)$ is the halo density in Equation \ref{eq:genDMDens}.  Thus, 
\begin{multline} 
    \psi(\infty) - \psi(\tilde{m}(\mathbf{x}, \tau)) = 2 \int \limits_0 ^ {\infty} d m' m'
     \begin{cases}
        \rho_{\textrm{in}} (m') & m' \leq c\ r_s \\
        0 & m'  > c\ r_s
    \end{cases}
    -  2 \int \limits_0 ^ {\sqrt{R_\tau^2 + z_\tau^2}} d m' m'
    \begin{cases}
        \rho_{\textrm{in}} (m') & m' \leq c \ r_s \\
        0 & m' > c \ r_s 
    \end{cases} \\
    = 2 \int \limits_0 ^ {c \ r_s} d m' m' \rho_{\textrm{in}} (m') - 2
    \begin{cases}
         \int \limits_0 ^ {\sqrt{R_\tau^2 + z_\tau^2}} d m' m' \rho_{\textrm{in}} (m') & \tau \geq \tau_c \\
        \int \limits_0 ^ {c \ r_s} d m' m' \rho_{\textrm{in}} (m')  & \tau < \tau_c 
    \end{cases} 
    = 2 
    \begin{cases}
         \int \limits_{\sqrt{R_\tau^2 + z_\tau^2}}^ {c \ r_s} d m' m' \rho_{\textrm{in}} (m') & \tau \geq \tau_c \\
        0 & \tau < \tau_c \ ,
    \end{cases} 
\end{multline} 
where $\tau_c$ is the solution to the equation 
\begin{equation} \label{eq:taucDef} 
\frac{R^2}{1 + \tau_c} + \frac{z^2}{Q^2 + \tau_c} \equiv (r_s c)^2 \ ,
\end{equation} 
and where we have defined $R_\tau^2 \equiv R^2/(1 + \tau)$ and $z_\tau^2 \equiv z^2/(Q^2 + \tau)$ for convenience. 

Equation \ref{eq:BandTElPhi} can thus be rewritten as 
\begin{multline}
    \Phi(\mathbf{x}) = - \pi G Q \int_0 ^ {\infty} d \tau \Big {(} \normalsize 2 
    \begin{cases}
         \int \limits_{\sqrt{R_\tau^2 + z_\tau^2}}^ {c \ r_s} d m' m' \rho_{\textrm{in}} (m') & \tau \geq \tau_c \\
        0 & \tau < \tau_c 
    \end{cases}      \Big ) \frac{1}{\sqrt{(\tau + 1)^2(\tau + Q^2)}} \\ 
    = - 2 \pi G Q r_s^2 \int_{\textrm{max}(\tau_c,0)} ^ {\infty} d \tau   \frac{1}{\sqrt{(\tau + 1)^2(\tau + Q^2)}}  \int \limits_{\sqrt{{R_\tau '}^2 + {z_\tau '}^2}}^ {c} d \mu' \mu' \rho_{\textrm{in}} (\mu') \ ,
\end{multline}
 where $\mu' \equiv m' / r_s$, ${R_\tau '} ^2 \equiv R^2/(r_s^2(1 + \tau))$, and ${z_\tau '} ^2 \equiv z^2/(r_s^2(Q^2 + \tau))$
 
 In this work, we consider three DM halo models: a `cusped' NFW halo, a `cored' Burkert halo, and a `cored' Acored halo.  Their interior DM densities are given by
\begin{equation} \label{eq:baseHalo}
    \rho_{\textrm{in}}  (\mu)= 
    \begin{cases}
       \frac{\rho_s}{\mu({1 + \mu})^2} & \textrm{(NFW)} \\
       \frac{\rho_s}{(1 + \mu)(1 + \mu ^2)} & \textrm{(Burkert)} \\ 
       \frac{\rho_s}{({1 + \mu})^3} & \textrm{(Acored)} \ . \\
    \end{cases} 
\end{equation}
We would rather express $\rho_{h}$ in terms of the total mass of the halo, $M_h$, defined as 
\begin{equation}
M_h \equiv \int_0 ^ {m_{cut}} dm \frac{dM_{h}(m)}{dm} = \int_0 ^ {m_{cut}} dm \rho_{h}(m) \frac{dV_{\textrm{spheroid}} }{dm} \ ,
\end{equation} 
where $V_{\textrm{spheroid}}  = 4/3\pi a^2 b$ is the volume of a spheroid of azimuthal radius $a$ and of symmetry axis length $b$.  If we consider the spheroid bounded by the constraint $\sqrt{a^2 + (b/Q)^2} = m$, then the volume as a function of $m$ is $V_{\textrm{spheroid}} (m) = 4/3 \pi m^2 Q m = 4/3 \pi Q m^3$.  Note that, if we did not impose the cutoff in Equation \ref{eq:genDMDens}, this definition of $M_h$ would diverge.  

We can now perform the integral for $M_h$ easily: 
\begin{equation}
M_h = \int_0 ^ {\infty} dm \rho(m)_{\textrm{DM},h} \frac{dV_{\textrm{spheroid}} }{dm} = \int_0 ^ {m_{cut}} dm  \rho_{i}(m)  4 \pi Q m^2 = \rho_s r_s^3 4 \pi Q \int_0 ^ {c} d\mu \frac{ \rho_{h}(\mu) }{\rho_s} \mu^2 \ ,
\end{equation} 
where we have defined $\mu = m /r_s$.

For the NFW halo, 
\begin{multline}
M_{h,\textrm{nfw}} = \rho_s r_s^3 4 \pi Q \int_0 ^ {c} d\mu \frac{1}{\mu({1 + \mu})^2} \mu^2 = \rho_s r_s^3 4 \pi Q \int_1 ^ {1+c} d\mu \frac{1}{{\nu}^2} (\nu -1 ) = \rho_s r_s^3 4 \pi Q [\ln{\nu} + \frac{1}{\nu}]_1 ^ {1+c} \\ 
= \rho_s r_s^3 4 \pi Q [\ln{(1+c)} - 0 + \frac{1}{1+c } - 1] = \rho_s r_s^3 4 \pi Q [\ln{(1+c)} -  \frac{c}{1+c }] \equiv \rho_s r_s^3 4 \pi Q f(c) \ ,
\end{multline} 
for the Acored halo,
\begin{multline}
M_{h,\textrm{Acored}} = \rho_s r_s^3 4 \pi Q \int_0 ^ {c} d\mu \frac{1}{({1 + \mu})^3} \mu^2 = \rho_s r_s^3 4 \pi Q \int_1 ^ {1+c} d\mu \frac{1}{{\nu}^3} (\nu -1 )^2 = \rho_s r_s^3 4 \pi Q [\ln{\nu} + \frac{2}{\nu} - \frac{1}{2 \nu^2}]_1 ^ {1+c} \\ 
= \rho_s r_s^3 4 \pi Q (\ln{(1+c)} + \frac{2}{1+c } - \frac{1}{2(1+c)^2} - 0 - 2 + \frac{1}{2}) = \rho_s r_s^3 4 \pi Q (\ln{(1+c)} -  \frac{2 c + 3 c^2}{2(1+c)^2 }) \equiv \rho_s r_s^3 4 \pi Q h(c) \ ,
\end{multline} 
and for the Burkert halo, 
\begin{multline}
M_{h,\textrm{Burkert}} = \rho_s r_s^3 4 \pi Q \int_0 ^ {c} d\mu \frac{\rho_s}{(1 + \mu)(1 + \mu ^2)} \mu^2 
= \rho_s r_s^3 4 \pi Q \int_0 ^ {0+c} d\mu \frac{1}{2} (\frac{1}{1+\mu} + \frac{\mu}{1 + \mu ^2} - \frac{1}{1+ \mu ^2} \\
= \rho_s r_s^3 4 \pi Q \frac{1}{2} [\ln{(1 + \mu)} + \frac{1}{2} \ln{(1 + \mu^2)} - \arctan{(\mu)}]_0 ^ c 
= (\rho_s r_s^3 4 \pi Q) \frac{1}{4} [2 \ln{(1 + c)} + \ln{(1 + c^2)} - 2 \arctan{(c)}] \\
\equiv (\rho_s r_s^3 4 \pi Q) g (c) \ .
\end{multline} 
So for all halos, we can express the mass density as a function of the halo mass: 
\begin{equation} \label{eq:hDensity}
    \rho_{\textrm{in}}  (R,z) = 
     \frac{M_h}{4 \pi Q r_s^3}
    \begin{cases}
            \frac{1}{f(c)} \frac{1}{\mu(R,z)} \frac{1}{(1+\mu(R,z))^2} & \textrm{(NFW)} \\
            \frac{1}{h(c)} \frac{1}{(1+\mu(R,z))^3} & \textrm{(Acored)}  \\
             \frac{1}{g(c)} \frac{1}{(1+\mu(R,z))(1 + \mu(R,z) ^ 2)} & \textrm{(Burkert)} \ ,
    \end{cases} 
\end{equation} 
where $f(c) \equiv \ln(1+c) -\frac{c}{1+c}$, $h(c) \equiv \ln{(1+c)} -  (2 c + 3 c^2)/(2(1+c)^2 ) $, and $g(c) = \frac{1}{4}(2 \ln{(1 + c)} + \ln{(1 + c^2)} - 2 \arctan{(c)} )$.  

We are now ready to compute the gravitational potential for each scenario:
\begin{multline} \label{eq:fullHaloPot} 
\Phi(\mathbf{x})    = - 2 \pi G Q r_s^2 \int_{\textrm{max}(\tau_c,0)} ^ {\infty} d \tau   \frac{1}{\sqrt{(\tau + 1)^2(\tau + Q^2)}}  \int \limits_{\sqrt{{R_\tau '}^2 + {z_\tau '}^2}}^ {c} d \mu' \mu' \frac{M_h}{4 \pi Q r_s^3}
    \begin{cases}
            \frac{1}{f(c)} \frac{1}{\mu'} \frac{1}{(1+\mu')^2} & \textrm{(NFW)} \\
            \frac{1}{h(c)} \frac{1}{(1+\mu')^3} & \textrm{(Acored)}  \\
             \frac{1}{g(c)} \frac{1}{(1+\mu')(1 + {\mu'} ^ 2)} & \textrm{(Burkert)}
    \end{cases} 
    \\
    =  \frac{- M_h G }{r_s} \int_{\textrm{max}(\tau_c,0)} ^ {\infty} d \tau   \frac{1}{\sqrt{(\tau + 1)^2(\tau + Q^2)}}  \int \limits_{\sqrt{{R_\tau '}^2 + {z_\tau '}^2}}^ {c} d \mu'
    \begin{cases}
            \frac{1}{2f(c)} \frac{1}{(1+\mu')^2} & \textrm{(NFW)} \\
            \frac{1}{2h(c)} \frac{\mu'}{(1+\mu')^3} & \textrm{(Acored)}  \\
             \frac{1}{2g(c)} \frac{\mu'}{(1+\mu')(1 + {\mu'} ^ 2)} & \textrm{(Burkert)}
    \end{cases} \\
    =  \frac{- M_h G }{r_s} \int_{\textrm{max}(\tau_c,0)} ^ {\infty} d \tau   \frac{1}{\sqrt{(\tau + 1)^2(\tau + Q^2)}} 
    \begin{cases}
         \frac{1}{2f(c)} \Big[\frac{-1}{1+\mu'} \Big] _{\mu' = {\sqrt{{R_\tau '}^2 + {z_\tau '}^2}}}^{\mu'=c}  & \textrm{(NFW)} \\
        \frac{1}{2h(c)}\Big[ \frac{-1}{1+\mu'} + \frac{1}{2} \frac{1}{(1+\mu')^2}\Big] _{\mu' = {\sqrt{{R_\tau '}^2 + {z_\tau '}^2}}}^{\mu'=c} & \textrm{(Acored)} \\
        \frac{1}{2g(c)} \Big [ \frac{1}{4} ( \ln{(\frac{1 + {\mu'}^2}{(1 + \mu')^2})} + 2\arctan{(\mu')} ) \Big] _{\mu' = {\sqrt{{R_\tau '}^2 + {z_\tau '}^2}}}^{\mu'=c} & \textrm{(Burkert)} 
    \end{cases} \\ 
    =  \frac{- M_h G }{r_s} \int_{\textrm{max}(\tau_c,0)} ^ {\infty} d \tau   \frac{1}{\sqrt{(\tau + 1)^2(\tau + Q^2)}} \\
    \times 
    \begin{cases}
         \frac{1}{2f(c)} \Big[\frac{-1}{1+c} + \frac{1}{1+{\sqrt{{R_\tau '}^2 + {z_\tau '}^2}}}\Big]  & \textrm{(NFW)} \\
        \frac{1}{2h(c)}\Big[ \frac{-1}{1+c} + \frac{1}{2} \frac{1}{(1+c)^2} + \frac{-1}{1+\sqrt{{R_\tau '}^2 + {z_\tau '}^2}} - \frac{1}{2} \frac{1}{(1+\sqrt{{R_\tau '}^2 + {z_\tau '}^2})^2}\Big] & \textrm{(Acored)}\\
        \frac{1}{8g(c)} \Big [ \ln{\frac{(1 + c^2) \Big( 1 + {\sqrt{{R_\tau '}^2 + {z_\tau '}^2}} \Big )^2} {(1 + c)^2 ({1 + {R_\tau '}^2 + {z_\tau '}^2})}} + 2\arctan{(c)}  - 2\arctan{(\sqrt{{R_\tau '}^2 + {z_\tau '}^2})} ) \Big ]& \textrm{(Burkert)} 
    \end{cases}
\end{multline} 
\begin{multline} \label{eq:fullHaloPot} 
\Rightarrow \Phi(\mathbf{x}, M_h, r_s, c, Q) =  \frac{- M_h G }{r_s} \int_{\textrm{max}(\tau_c,0)} ^ {\infty} d \tau   \frac{1}{\sqrt{(\tau + 1)^2(\tau + Q^2)}} \\
    \times 
    \begin{cases}
         \frac{1}{2f(c)} \Big[\frac{-1}{1+c} + \frac{1}{1+{\sqrt{{R_\tau '}^2 + {z_\tau '}^2}}}\Big]  & \textrm{(NFW)} \\
        \frac{1}{4h(c)}\Big[ -\frac{1+2c}{(1+c)^2} + \frac{1 + 2 \sqrt{{R_\tau '}^2 + {z_\tau '}^2}}{\Big (1+\sqrt{{R_\tau '}^2 + {z_\tau '}^2} \Big)^2} \Big] & \textrm{(Acored)} \\
        \frac{1}{8g(c)} \Big [ \ln{\frac{(1 + c^2) \Big( 1 + {\sqrt{{R_\tau '}^2 + {z_\tau '}^2}} \Big )^2} {(1 + c)^2 ({1 + {R_\tau '}^2 + {z_\tau '}^2})}} + 2\arctan{(c)}  - 2\arctan{(\sqrt{{R_\tau '}^2 + {z_\tau '}^2})} ) \Big ]& \textrm{(Burkert)} 
    \end{cases} \\
     \equiv \frac{- M_h G }{r_s} \Gamma (\frac{R}{r_s},\frac{z}{r_s},Q,c) \ ,
\end{multline} 
where recall $\tau_c$ is defined in equation  \ref{eq:taucDef}.  We have thus reduced the computation of the gravitational potential for an spheroidal NFW,  Acored, or Burkert halo to the computation of the single numerical integral in Equation \ref{eq:fullHaloPot} for particular values of $R/r_s$ and $z/r_s$ (the halo coordinates scaled by the halo scale radius), $c$ (the number of scale radii at which we truncate the halo mass), and $Q$ (the ellipticity of the halo). 

\section{Computing $\Phi_{d}$} \label{app:diskPot} 

Here, we determine the potential sourced by an exponential disk with density distribution of form
\beq \label{eq:sechDisk}
\rho_D(R,z) = \frac{M_d}{8\pi R_d^2 z_d} \exp{\Big (-\frac{R}{R_d} \Big )} \sech^2 \left( \frac{z}{2z_d} \right)  
\eeq
using the methodology outlined in Chapter 3 of \citep{BandT}. 
Note that
\beq
\int dz dR^2 \rho_D(R,z) = M_d \ .
\eeq

We will start by considering the scenario of an infinitely thin disk.  To determine the gravitational potential from a mass density distribution, we must solve the Poisson equation in cylindrical coordinates:
\beq \label{eq:cylPoisson}
\frac{1}{R} \frac{\partial}{\partial R} \left( R \frac{\partial \Phi}{\partial R}\right) + \frac{1}{R^2} \frac{\partial^2 \Phi}{\partial \psi^2} + \frac{\partial^2 \Phi}{\partial z^2} = \rho \ . 
\eeq
We first solve for the homogenous solution ($\rho = 0$) by separating variables.  Specifically, we define $\Phi \equiv J(R) F(\psi) Z(z)$ and note that each term in the Equation \ref{eq:cylPoisson} depends on only one variable and thus must independently vanish when $\rho = 0$.  Therefore,  
\begin{align}
0=& \frac{d^2 Z}{dz^2} - k^2 Z  & \Rightarrow &Z(z) \propto e^{\pm kz} \ , \\
0=&\frac{d^2 F}{d\psi^2} + m^2 F& \Rightarrow &F(\psi) \propto e^{im\psi} \ , \\
0=& R \frac{d}{dR} \left(R \frac{dJ}{dR}\right) + k^2 R^2 J(R) - m^2 J & \Rightarrow &J(R) \propto J_m(k R) \ , 
\end{align}
where $m$ is an integer (periodic boundary conditions), $k$ is a real number, and $J_m$ are the Bessel functions.  Recall that the Bessel functions obey the orthogonality relation: 
\beq \label{eq:BesselOrtho}
\int J_\nu(kr) J_\nu(k'r) r dr = \frac{\delta(k-k')}{k} \ .
\eeq
We thus find that an space with no matter permits a potential of the form 
\beq \label{eq:thinkDiskPot} 
\Phi_{k,m} = Ae^{im\psi - k|z|} J_m(kR)
\eeq
for some constant A.  This function has a discontinuity in the first derivative at $z=0$:
\begin{align}
\lim_{z\to0_\pm} = A \mp k \Phi(R,0) \ ,
\end{align}
and can thus be identified as the potential of an infinitely thin disk with surface density
\beq
\Sigma_{k,m}(R,\psi) = -A \frac{k}{2\pi G} e^{im\psi} J_m(k R)
\eeq
(note the sign is important, since a positive mass density demands a negative $A$, and thus a negative surrounding potential).  

The rest of the problem is based on the superposition principle. If we can find a function $S_m(k)$ such that
\beq
\Sigma(R) = \sum_{m=-\infty}^{m=\infty} \int dk S_m(k) \Sigma_{k,m} = -\sum_{m=-\infty}^{m=\infty} \int dk S_m(k)\frac{k}{2\pi G} e^{im\psi} J_m(k R),
\eeq
then the potential for an infinitely thin disk with surface density $\Sigma(R)$ is
\beq
\Phi = \sum_{m=-\infty}^{m=\infty} \int dk S_m(k) e^{im\psi-k|z|} J_m(k R) 
\eeq
(note that we have subsumed the general constant $A$ into the general function $S_m(k)$).  Because we are interested in axisymmetric solutions, we will search for systems with $m=0$.  We now invert the expression 
\beq
\Sigma(R) = -\int dk S_0(k)\frac{k}{2\pi G} J_0(k R)
\eeq
to get $S_0(k)$ for a given $\Sigma(R)$. Multiplying both sides by $R J_0(R \kappa)$, integrating over $R$, and using the orthogonality of the Bessel functions (Equation \ref{eq:BesselOrtho}), we find 
\beq
\int dR R \Sigma(R)J_0(\kappa R) = & -\int R dR dk S_0(k)\frac{k}{2\pi G} J_0(k R) J_0(\kappa R) \\
                                                          &= -\int dR R dk S_0(k)\frac{k}{2\pi G} J_0(k R) J_0(\kappa R)\\
                                                          &= -\int dk S_0(k)\frac{k}{2\pi G} \frac{\delta(k-\kappa)}{k}\\
                                                          &= -\int dk S_0(k)\frac{1}{2\pi G} \delta(k-\kappa)\\
                                                          &= - \frac{S_0(\kappa)}{2\pi G} \ .
\eeq
We can thus relate the potential for an infinitely thin disk to the surface density of the infinitely thin disk via 
\beq
\Phi(R,z) = -2\pi G\int dk e^{-k|z|} J_0(k R) \int dR' R' \Sigma(R')J_0(k R').
\eeq

A thick disk is a superposition of many (stacked) thin disks.  Each thin disk is at a different height along the $z$ axis with a surface mass density $\Sigma(R,z) = dz \rho(R,z)$. Thus, a thick disk has a potential 
\beq \label{eq:thinkDiskPot}
\Phi(R,z) = \int d\Phi = -2\pi G \int d z' \int dkJ_0(k R)   e^{-k|z - z' |}  \int dR' R' \rho(R',z')J_0(k R') \ .
\eeq

Hence, the potential for the thick disk in Equation \ref{eq:sechDisk} is:
\beq
\Phi(R,z) & = \int d\Phi  = -\frac{ GM_d}{ 4 R_d^2 z_d } \int dkJ_0(k R)\int dz'  e^{-k|z-z'|} \sech^2 \left( \frac{z'}{2z_d} \right)  \int dR' R'  \exp(-R'/R_d)J_0(k R') \ .
\eeq
The radial integral can be easily solved:
\beq
\int dR' R'  \exp(-R'/R_d)J_0(k R') = \frac{R_d^2}{(1+k^2 R_d^2)^{3/2}} \ . 
\eeq
The remaining integrals can be expressed in terms of dimensionless coordinates, $\alpha \equiv R/R_d$ and $\beta \equiv z/z_d$, and the dimensionless morphology parameter $\lambda \equiv z_d / R_z$: 
\beq \label{eq;PhiFroFirstIntegral}
\Phi(\alpha,\beta,\lambda) &= -\frac{ GM_d}{ 4 z_d }  \int  \frac{dk J_0(k R_d \alpha)}{(1+k^2 R_d^2)^{3/2}} \int dz'  e^{-k|\beta z_d-z'|} \sech^2 \left( \frac{z'}{2z_d} \right) \\
&= -\frac{ GM_d}{ 4 R_d }  \int  \frac{d(k R_d) J_0((k R_d) \alpha)}{(1+(k R_d)^2)^{3/2}} \int d(z'/z_d)  e^{-(k R_d) (z_d / R_d) |\beta-(z' / z_d)|} \sech^2 \left( \frac{(z'/z_d)}{2} \right) \\
& = -\frac{ GM_d}{ 4R_d }  \int  \frac{d\kappa J_0(\kappa \alpha)}{(1+\kappa^2)^{3/2}} \int_{-\infty}^{\infty} d\zeta  e^{-\kappa \lambda|\beta -\zeta|} \sech^2 \left( \frac{\zeta}{2} \right) \\
&= \int  \frac{d\kappa J_0(\kappa \alpha)}{(1+\kappa^2)^{3/2}} I (\kappa, \beta, \lambda) \ . 
\eeq

We now give an analytic expression of the $\zeta$ integral:
\begin{equation}
I (\kappa, \beta, \lambda) = -\frac{G M_d}{4 z_d} \int_{-\infty}^{\infty}d\zeta e^{- \kappa \lambda |\beta -\zeta|} \textnormal{ sech}^2 (\zeta/2) \ . 
\end{equation}

To do this computation, we first separate the single integral into two integrals to handle the absolute value:
\begin{equation}
\int_{-\infty}^{\infty}d\zeta e^{- \kappa \lambda |\beta -\zeta|}\textnormal{ sech}^2 (\zeta/2) 
= \int_{-\infty}^{\beta}d\zeta e^{- \kappa \lambda (\beta -\zeta)}\textnormal{ sech}^2 (\zeta/2)  +  \int_{\beta}^{\infty}d\zeta e^{- \kappa \lambda (\zeta-\beta)} \textnormal{ sech}^2 (\zeta/2) \ , 
\end{equation}
and then pull out variables independent of $\zeta$ and express the integral in terms of $\gamma \equiv \zeta/2$:
\begin{multline}
 \int_{-\infty}^{\beta}d\zeta e^{- \kappa \lambda (\beta -\zeta)}\textnormal{ sech}^2 (\zeta/2) +  \int_{\beta}^{\infty}d\zeta e^{- \kappa \lambda (\zeta-\beta)}\textnormal{ sech}^2 (\zeta/2) \\
=  2e^{-\kappa  \lambda \beta }\int_{-\infty}^{\beta/2}d\gamma e^{2 \kappa \lambda \gamma } \textnormal{ sech}^2 (\gamma) + 2e^{\kappa  \lambda \beta } \int_{\beta/2}^{\infty}d\gamma e^{- 2 \kappa \lambda \gamma}\textnormal{ sech}^2 (\gamma) \ . 
\end{multline}
Now we integrate by parts, noting that $d (\tanh {\gamma})/d\gamma =\textnormal{ sech}^2{\gamma}$: 
\begin{multline}
2e^{-\kappa  \lambda \beta }\int_{-\infty}^{\beta/2}d\gamma e^{2 \kappa \lambda \gamma } \textnormal{ sech}^2 (\gamma)  + 2e^{\kappa  \lambda \beta} \int_{\beta/2}^{\infty}d\gamma e^{- 2 \kappa \lambda \gamma}\textnormal{ sech}^2 (\gamma) \\
=2e^{-\kappa \lambda \beta} \{ [ e^{2 \kappa \lambda \gamma} \tanh (\gamma)]_{\gamma=-\infty}^{\gamma=\beta/2} - \int_{-\infty}^{\beta/2}d\gamma 2 \kappa \lambda  e^{2 \kappa \lambda \gamma } \tanh (\gamma) \} \\ + 2e^{\kappa \lambda \beta} \{ [ e^{-2 \kappa \lambda \gamma} \tanh (\gamma)]_{\gamma=\beta/2}^{\gamma=\infty} - \int_{\beta/2}^{\infty}d\gamma (-2 \kappa \lambda)  e^{-2 \kappa \lambda \gamma } \tanh (\gamma) \} \\
=2e^{-\kappa \lambda \beta} \{ [e^{\kappa \lambda \beta} \tanh (\beta/2)-e^{-\infty} \tanh (-\infty)] - 2 \kappa \lambda \int_{-\infty}^{\beta/2}d\gamma  (e^{\gamma}e^{-\gamma}) (e^{\gamma })^{2 \kappa \lambda } \frac{e^{\gamma} - e^{-\gamma}}{e^{\gamma} + e^{-\gamma}} \}  \\
+ 2e^{\kappa \lambda \beta} \{ [e^{ - \infty} \tanh (\infty)-e^{-\kappa \beta \lambda} \tanh (\beta/2)] + 2 \kappa \lambda \int_{\beta/2}^{\infty}d\gamma  (e^{\gamma}e^{-\gamma})(-1)(-1) (e^{-\gamma })^{2 \kappa \lambda } \frac{e^{\gamma} - e^{-\gamma}}{e^{\gamma} + e^{-\gamma}} \} \ . 
\end{multline}
Redefining integration variables $ u \equiv e^{\gamma}$ and $v \equiv e^{-\gamma}$  $\Rightarrow$ $du =d \gamma  e^{\gamma} $ and $dv =-d \gamma  e^{-\gamma} $, we continue:  
\begin{multline}
2e^{-\kappa \lambda \beta} \{ [e^{\kappa \lambda \beta} \tanh (\beta/2)-e^{-\infty} \tanh (-\infty)] - 2 \kappa \lambda \int_{-\infty}^{\beta/2}d\gamma  (e^{\gamma}e^{-\gamma}) (e^{\gamma })^{2 \kappa \lambda } \frac{e^{\gamma} - e^{-\gamma}}{e^{\gamma} + e^{-\gamma}} \}  \\
+ 2e^{\kappa \lambda \beta} \{ [e^{ - \infty} \tanh (\infty)-e^{-\kappa \beta \lambda} \tanh (\beta/2)] + 2 \kappa \lambda \int_{\beta/2}^{\infty}d\gamma  (e^{\gamma}e^{-\gamma})(-1)(-1) (e^{-\gamma })^{2 \kappa \lambda } \frac{e^{\gamma} - e^{-\gamma}}{e^{\gamma} + e^{-\gamma}} \}  \\
=2e^{-\kappa \lambda \beta} \{ [e^{\kappa \lambda \beta} \tanh (\beta/2) - 0] - 2 \kappa \lambda \int_{e^{-\infty}}^{e^{\beta/2}}du ( u^{-1}) (u)^{2 \kappa \lambda } \frac{u - u^{-1}}{u +u^{-1}} \}  \\
+ 2e^{\kappa \lambda \beta} \{ [0-e^{-\kappa \beta \lambda} \tanh (\beta/2)] - 2 \kappa \lambda \int_{e^{-\beta/2}}^{e^{-\infty}}dv  v^{-1} (v)^{2 \kappa \lambda } \frac{v^{-1} - v}{v^{-1} +v} \}  \\
= 2 \tanh (\beta/2)  - 4 \kappa \lambda e^{-\kappa \lambda \beta} \int_{e^{-\infty}}^{e^{\beta/2}}du  \ u^{2 \kappa \lambda } \frac{1 - u^{-2}}{u +u^{-1}} -2 \tanh (\beta/2) + 4 \kappa \lambda e^{\kappa \lambda \beta} \int_{e^{-\infty}}^{e^{-\beta/2}}dv \ v^{2 \kappa \lambda } \frac{v^{-2} - 1}{v^{-1} +v}   \\
= 4 \kappa \lambda \ \Big(-  e^{-\kappa \lambda \beta} \int_{e^{-\infty}}^{e^{\beta/2}}du  \ u^{2 \kappa \lambda } \frac{1 - u^{-2}}{u +u^{-1}} -  e^{\kappa \lambda \beta} \int_{e^{-\infty}}^{e^{-\beta/2}}dv \ v^{2 \kappa \lambda } \frac{1- v^{-2} }{v^{-1} +v} \Big) \ ,
\end{multline}
 where we used the fact that $e^{-\infty} \tanh{(\pm \infty)}$ = 0
According to the definition of the hypergeometric function, $_2 F _1 (a,b;c;d)$:
\begin{equation}
\int dx \ x^{2k} \frac{1-x^{-2}}{x+x^{-1}} = \frac{1}{2} x^{2k} \Big(\frac{2 x^2 \ _2 F _1 (1,k +1; k+2 ;-x^2)}{k+1} - \frac{1}{k} \Big) + (\textrm{constant}) \ .
\end{equation}
So we find:
\begin{multline}
4 \kappa \lambda \ \Big(- e^{-\kappa \lambda \beta} \int_{0}^{e^{\beta/2}}du  \ u^{2 \kappa \lambda } \frac{1 - u^{-2}}{u +u^{-1}} - e^{\kappa \lambda \beta} \int_{0}^{e^{-\beta/2}}dv \ v^{2 \kappa \lambda } \frac{1- v^{-2} }{v^{-1} +v} \Big) \\
= 4 \kappa \lambda \Big( - e^{-\kappa \lambda \beta} \Big[ \frac{1}{2} u^{2\kappa \lambda} \Big(\frac{2 u^2 \ _2 F _1 (1,\kappa \lambda +1; \kappa \lambda+2 ;-u^2)}{\kappa \lambda+1} - \frac{1}{\kappa \lambda}\Big) \Big]_{0}^{e^{\beta/2}} \\
- \ e^{\kappa \lambda \beta} \Big[ \frac{1}{2} v^{2\kappa \lambda } \Big(\frac{2 v^2 \ _2 F _1 (1,\kappa \lambda +1; \kappa \lambda +2 ;-v^2)}{\kappa \lambda +1} - \frac{1}{\kappa \lambda}\Big) \Big]_{0}^{e^{-\beta/2}} \Big) \\ 
= 2 \kappa \lambda \Big( - e^{-\kappa \lambda \beta} \Big[ (e^{\beta/2})^{2\kappa \lambda } \Big(\frac{2  (e^{\beta/2})^2 \ _2 F _1 (1,\kappa \lambda  +1; \kappa \lambda +2 ;- (e^{\beta/2})^2)}{\kappa \lambda +1} - \frac{1}{\kappa \lambda }\Big) -(-0) \Big]\\
 -  e^{\kappa \lambda \beta}  \Big[ (e^{-\beta/2})^{2\kappa \lambda } \Big(\frac{2  (e^{-\beta/2})^2 \ _2 F _1 (1,\kappa \lambda  +1; \kappa \lambda +2 ;- (e^{-\beta/2})^2)}{\kappa \lambda +1} - \frac{1}{\kappa \lambda }\Big) -(-0) \Big] \Big) \\ 
= 2 \kappa \lambda \Big[ - e^{-\kappa \lambda \beta} \ e^{\beta \kappa \lambda } \Big(\frac{2  e^{\beta} \ _2 F _1 (1,\kappa \lambda  +1; \kappa \lambda +2 ;- e^{\beta})}{\kappa \lambda +1} - \frac{1}{\kappa \lambda }\Big)\\
 -  e^{\kappa \lambda \beta}  e^{-\beta \kappa \lambda } \Big(\frac{2  e^{-\beta} \ _2 F _1 (1,\kappa \lambda  +1; \kappa \lambda +2 ;- e^{-\beta})}{\kappa \lambda +1} - \frac{1}{\kappa \lambda }\Big) \Big]  \\ 
=  -\frac{4 \kappa \lambda  e^{\beta} \ _2 F _1 (1,\kappa \lambda  +1; \kappa \lambda +2 ;- e^{\beta})}{\kappa \lambda +1} + 2
 -  \frac{4 \kappa \lambda  e^{-\beta} \ _2 F _1 (1,\kappa \lambda  +1; \kappa \lambda +2 ;- e^{-\beta})}{\kappa \lambda +1} +2 \\
= 4 \Big(1 - \frac{\kappa \lambda  e^{\beta} \ _2 F _1 (1,\kappa \lambda  +1; \kappa \lambda +2 ;- e^{\beta})}{\kappa \lambda +1} - \frac{\kappa \lambda  e^{-\beta} \ _2 F _1 (1,\kappa \lambda  +1; \kappa \lambda +2 ;- e^{-\beta})}{\kappa \lambda +1}\Big) \ .
\end{multline}
Thus, we finally determine that 
\begin{multline}
\frac{G M_d}{4 z_d} \int_{-\infty}^{\infty}d\zeta e^{- \kappa \lambda |\beta -\zeta|} \textnormal{ sech}^2 (\zeta/2) \\ 
= \frac{G M_d}{z_d} \Big (1 - \frac{\kappa \lambda}{\kappa \lambda +1}(e^{\beta}\ _2 F _1 (1,\kappa \lambda  +1; \kappa \lambda +2 ;- e^{\beta}) + e^{-\beta}\ _2 F _1 (1,\kappa \lambda  +1; \kappa \lambda +2 ;- e^{-\beta}) ) \Big) \ .
\end{multline}

Inserting this result into Equation \ref{eq;PhiFroFirstIntegral}, we express $\Phi_{d}$ as a single integral:
\begin{multline} \label{eq:fullDiskPot} 
\Phi_{d} (\alpha, \beta, M_d, R_d, \lambda) = - \frac{G M_d}{z_d}  \int d \kappa \frac{J_0 (\kappa \alpha)} {(1+ \kappa ^2)^{3/2}}  \\ 
\times \Big (1 - \frac{\kappa \lambda}{\kappa \lambda +1}(e^{\beta}\ _2 F _1 (1,\kappa \lambda  +1; \kappa \lambda +2 ;- e^{\beta}) + e^{-\beta}\ _2 F _1 (1,\kappa \lambda  +1; \kappa \lambda +2 ;- e^{-\beta}) ) \Big) \\ 
\equiv -\frac{G M_d}{R_d \lambda} F(\alpha,\beta,\lambda) \ ,
\end{multline} 
where $\alpha$ and $\beta$ are related to the dimensionfull coordinates, R and z by $\alpha = R/R_d $ and $\beta = z/(\zeta R_d)$.


\end{document}